\documentclass[11pt,a4paper]{article}
\usepackage{jheppub}

\usepackage{amsmath}
\usepackage{amsfonts}
\usepackage{slashed}
\usepackage{graphicx}
\usepackage{mathrsfs}
\usepackage{tikz}
\usepackage[numbers,sort&compress]{natbib}
\usepackage{hyperref}
\usepackage{xcolor}
\usepackage{verbatim}
\usepackage[backgroundcolor=white,linecolor=black]{todonotes}

\makeatletter
\g@addto@macro\bfseries{\boldmath}
\makeatother

\newcommand{\eps}{\epsilon}
\newcommand{\ord}{\begin{cal}O\end{cal}}

\def\beq{\begin{equation}}
\def\eeq{\end{equation}}
\def\bsp#1\esp{\begin{split}#1\end{split}}

\makeatletter
\newcommand{\IEIF}{%
  \def\@IEIFsep{(}%
  I_F\@IEIFi
}
\newcommand\@IEIFi{\@ifnextchar\stopIEIF{\@IEIFend}{\@IEIFii}}
\newcommand\@IEIFii[4]{%
  \big\@IEIFsep
  \begin{smallmatrix}
    #1 & #2 \\
    #3 & #4
  \end{smallmatrix}
  \def\@IEIFsep{|}
  \@IEIFi
}
\newcommand\@IEIFend[2]{%
  ; #2 \bigr)
}
\makeatother

\newcommand{\cC}{\begin{cal}C\end{cal}}

\newcommand{\cZ}{\begin{cal}Z\end{cal}}

\newcommand{\rpp}{1-r_{+}}
\newcommand{\rpm}{1-r_{-}}
\newcommand{\rmp}{r_{+}}
\newcommand{\rmm}{r_{-}}

\newcommand{\Gm}{G_{-}}
\newcommand{\Rp}{R_{+}}
\newcommand{\Rm}{R_{-}}

\newcommand{\gamtt}[4]{{\widetilde{\Gamma}}\!\left(\begin{smallmatrix}#1\\#2\end{smallmatrix};#3,#4\right)}

\newcommand{\cEfe}[4]{{\mathcal{E}_4}\!\left(\begin{smallmatrix}#1\\#2\end{smallmatrix};#3,#4\right)}
\newcommand{\cEf}[3]{{\mathcal{E}_4}\!\left(\begin{smallmatrix}#1\\#2\end{smallmatrix};#3\right)}

\newcommand{\GG}{G\!}

\renewcommand{\ln}{\log}

\allowdisplaybreaks

\title{Elliptic polylogarithms and Feynman parameter integrals}

\author[a]{Johannes Broedel} 
\author[b,c]{Claude Duhr}
\author[d]{Falko Dulat}
\author[b]{Brenda Penante}
\author[b]{Lorenzo Tancredi}

\affiliation[a]{Institut f\"{u}r Mathematik und Institut f\"{u}r Physik,
Humboldt-Universit\"{a}t zu Berlin,\\
IRIS Adlershof, Zum Grossen Windkanal 6, 12489 Berlin, Germany} 
\affiliation[b]{Theoretical Physics Department, CERN, Geneva, Switzerland} 
\affiliation[c]{Center for Cosmology, Particle Physics and Phenomenology (CP3),\\
Universit\'e Catholique de Louvain, 1348 Louvain-La-Neuve, Belgium}
\affiliation[d]{SLAC National Accelerator Laboratory, Stanford University, Stanford, CA 94309, USA}

\emailAdd{jbroedel@physik.hu-berlin.de}
\emailAdd{claude.duhr@cern.ch}
\emailAdd{dulatf@slac.stanford.edu}
\emailAdd{b.penante@cern.ch}
\emailAdd{lorenzo.tancredi@cern.ch}

\abstract{
In this paper we study the calculation of multiloop Feynman integrals that cannot be expressed in terms of multiple polylogarithms. 
We show in detail how certain types of two- and three-point functions at two loops, which 
appear in the calculation of higher order corrections in QED, QCD and in the electroweak theory (EW), 
can naturally be expressed  in terms of a recently introduced elliptic generalisation of multiple polylogarithms by direct 
integration over their Feynman parameter representation.
Moreover, we show that in all examples that we considered a basis of pure Feynman integrals can be found.
}

\preprint{\begin{minipage}[t]{8cm}\begin{flushright}CP3-19-07, CERN-TH-2019-016\\
            HU-Mathematik-2019-01, HU-EP-19/03\\
            SLAC-PUB-17406\end{flushright}\end{minipage}}

\begin{document}

\maketitle

\catcode`\@=11
\font\manfnt=manfnt
\def\Watchout{\@ifnextchar [{\W@tchout}{\W@tchout[1]}}
\def\W@tchout[#1]{{\manfnt\@tempcnta#1\relax%
  \@whilenum\@tempcnta>\z@\do{%
    \char"7F\hskip 0.3em\advance\@tempcnta\m@ne}}}
\let\foo\W@tchout
\def\dubious{\@ifnextchar[{\@dubious}{\@dubious[1]}}
\let\enddubious\endlist
\def\@dubious[#1]{%
  \setbox\@tempboxa\hbox{\@W@tchout#1}
  \@tempdima\wd\@tempboxa
  \list{}{\leftmargin\@tempdima}\item[\hbox to 0pt{\hss\@W@tchout#1}]}
\def\@W@tchout#1{\W@tchout[#1]}
\catcode`\@=12


\section{Introduction}
\label{sec:intro}

Feynman integrals constitute the building blocks for the 
study of scattering processes in perturbative quantum field theory (QFT).
High precision calculations in QFT require the ability to compute increasingly more complicated Feynman integrals which 
involve many internal loops and external legs.
The success of the collider physics program, highlighted in the last years by the impressive results obtained by the LHC at CERN,
has pushed the required precision of theoretical computations to an unprecedented level. In order to keep up with the experimental demands, theoretical predictions of processes involving Feynman 
integrals with internal masses and with at least two loops and up to five external legs have become mandatory.
In spite of the extreme complexity of these calculations, the last two decades have witnessed an impressive
advancement in our understanding of perturbative QFT and, as a result, of our ability to
keep such calculations under control.

Typically, complicated Feynman integrals are computed by means of two seemingly orthogonal methods.
On the one hand, one can attempt their direct integration over some integral representation (for example in terms
of Feynman parameters or Mellin-Barnes integrals). On the other hand, one can derive differential equations (DE)
satisfied by the Feynman
integrals and try to solve them~\cite{Kotikov:1990kg,Bern:1993kr,Remiddi:1997ny,Gehrmann:1999as}.
Understanding the importance of multiple polylogarithms (MPLs)~\cite{Kummer,Nielsen,Goncharov:1998kja} 
in high-energy physics~\cite{Remiddi:1999ew,Gehrmann:2000zt}, and the study of their analytical, 
algebraic~\cite{GoncharovMixedTate,Duhr:2011zq,Duhr:2012fh} and numerical~\cite{Vollinga:2004sn} properties, 
have been crucial steps to systematise both strategies.
For what concerns direct integration techniques, this program culminated in the 
enunciation of the criterion of linear reducibility~\cite{Brown:2008um,Panzer:2015ida}, which 
allows one to define a 
(quite general) class of Feynman integrals that can be algorithmically expressed in terms of MPLs
by direct integration over their Feynman-parameter representation.
A similarly important result in the context of the differential-equation method 
(even if mathematically less well established) is the concept of 
canonical basis of master integrals~\cite{Henn:2013pwa}. Canonical master integrals
fulfil differential equations which admit solutions in terms of iterated integrals 
over particularly simple kernels, which in turn can be expressed as total differentials of logarithms.
If there exists a parametrisation of the external kinematics in terms of which the 
arguments of the logarithms are all rational functions,
the corresponding iterated integrals can be straightforwardly expressed in terms of MPLs. 
Indeed, both by direct integration through the linear reducibility criterion, and in the case of canonical differential 
equations\footnote{In the absence of square roots which cannot be rationalised.},  
one ends up with iterated integrals in a set of variables over rational functions in these variables. It is then 
a well-known fact that the space defined by these integrals 
is spanned by linear combinations of MPLs and rational functions.

Despite their applicability to large classes of problems in high-energy physics, already at the second loop order
MPLs are known not to exhaust the whole space 
of special functions required for the computation of Feynman integrals. 
Indeed, as early as 1962  A. Sabry, in an attempt to compute the two-loop 
corrections to the electron propagator in QED, encountered integrals of complicated algebraic functions 
which could not be evaluated in terms of polylogarithms~\cite{Sabry}, but instead required the introduction of
elliptic integrals and integrals thereof.
It was not until the second decade of the twenty-first century that such integrals came back to the centre of investigation
in particle physics, when it was realised that similar mathematical objects were required for the computation of 
multiloop corrections to processes of crucial importance to the physics programme at the LHC, like the production of $t\bar{t}$ pairs in NNLO QCD.
Since then, 
the development of techniques to treat integrals beyond MPLs has been a very pressing issue, both for their potential phenomenological impact in collider physics,
and also for their conceptual relevance~\cite{Broadhurst:1987ei,Bauberger:1994by,Bauberger:1994hx,Laporta:2004rb,Kniehl:2005bc,Aglietti:2007as,Czakon:2008ii,BrownLevin,Bloch:2013tra,Adams:2013nia,Adams:2014vja,Adams:2015gva,Adams:2015ydq,Remiddi:2016gno,Primo:2016ebd,Bonciani:2016qxi,Bloch:2016izu,Adams:2016xah,Passarino:2016zcd,vonManteuffel:2017hms,Primo:2017ipr,Remiddi:2017har,Broedel:2017kkb,Ablinger:2017bjx,Chen:2017pyi,Hidding:2017jkk,Broedel:2018iwv,Adams:2018bsn,Adams:2018kez,Broedel:2018rwm,Adams:2018ulb,Blumlein:2018aeq,Blumlein:2018jgc,Vanhove:2018mto}.
Thanks to this concerted effort, in the last years significant steps have been taken in extending both strategies (i.e.~direct integration and differential equations)
to the so-called elliptic case, namely when the natural geometry associated to the Feynman graphs under consideration is related to a family of Riemann surfaces of genus one.

The scope of this paper is to show how the algorithms for the direct integration of Feynman integrals in terms of MPLs 
can be suitably generalised to the elliptic case
by exploiting the properties of the elliptic polylogarithms (eMPLs)
defined in refs.~\cite{Broedel:2018qkq,Broedel:2017siw,Broedel:2017kkb}.
We recall here that these eMPLs are essentially equivalent to the multiple elliptic polylogarithms 
defined by Brown and Levin in ref.~\cite{BrownLevin}. 
In particular, by working out different examples explicitly, we demonstrate how to treat those classes of Feynman integrals which do not 
fulfil the criterion of linear reducibility and, instead,
require dealing with iterated integrals over more general
rational functions $R(x,y)$, where $y=\sqrt{P(x)}$ defines an elliptic curve\footnote{In this case, $P(x)$ can be an irreducible cubic or quartic polynomial.}.
We stress that, in order for our approach to be successful, one must deal with integrals where only one single elliptic curve
appears and no other square roots are present. 
In this sense, this paper constitutes a concrete step towards the generalisation of the machinery 
developed to integrate Feynman integrals in terms of MPLs to the elliptic case. 

As an important by-product of our calculations, we show that in all examples that we have 
considered, a basis of \textsl{pure master integrals} can be defined, following the definition provided in ref.~\cite{Broedel:2018qkq}. 
The existence of a pure basis of master integrals is conjectured in the polylogarithmic case and the results presented in this 
paper constitute non-trivial evidence of a possible generalisation of the current conjectures to Feynman integrals beyond MPLs.

It is possible to draw a parallel between our approach and the idea of ``elliptic linear reducibility'' recently put forward in ref.~\cite{Hidding:2017jkk}.
There, a generalisation of the criterion of linear reducibility to the elliptic case is attempted.
Our approach is different since we work in the framework of a well defined class of functions whose 
algebraic and analytic properties can be studied rigorously and for which a concept of \textsl{transcendental weight}
can be defined. 
 
The possibility of evaluating Feynman integrals in terms of a well-known class of functions 
with understood algebraic properties is not only important for computational reasons, 
but can also be of conceptual relevance. Indeed, the notion of transcendental weight, which is an integer number associated to a pure function, can be used as an organisational 
tool to classify different expressions. For specific theories, such as $\mathcal{N}=4$ super Yang-Mills, 
scattering amplitudes at every loop order $L$ are believed to be of strictly maximal weight $2L$. 
This property is obeyed by all known examples and can significantly reduce the space of 
functions needed to represent an amplitude. This fact has been heavily explored by the 
amplitudes bootstrap community in refs.~\cite{Dixon:2011nj,Dixon:2011pw,Dixon:2013eka,Dixon:2014iba,Dixon:2014voa,Drummond:2014ffa,Dixon:2015iva,Caron-Huot:2016owq,Dixon:2016nkn} in order to obtain results up to four loops and seven external legs or five loops and six external legs.
Although every amplitude in $\mathcal{N}=4$ super Yang-Mills which evaluates to MPLs 
is indeed of uniform maximal weight, it is known that more complicated functions (of the 
elliptic kind and beyond) are inevitable also in this theory starting already at two loops~\cite{CaronHuot:2012ab,Bourjaily:2018yfy}. 
Therefore, extending the notion of transcendental weight to functions beyond MPLs is a crucial step in order 
to test the conjecture that observables in $\mathcal{N}=4$ super Yang-Mills evaluate to functions of uniform weight. 

Before diving into the computations, one more comment is in order.
It is very clear to us that, even at two loops, the class of functions that we are considering will probably not be the end of the story, since either 
multiple elliptic curves~\cite{Adams:2018bsn,Adams:2018kez},
or entirely new geometrical objects~\cite{Bloch:2014qca,Primo:2017ipr,Bourjaily:2017bsb,Bourjaily:2018ycu,Bourjaily:2018yfy} can appear. 
Still, with this paper we aim to show that our
framework is general and flexible enough to cover many problems of direct physical interest and, therefore, deserves
to be developed further.
We will show explicitly how different  two- and three-point functions at two loops can be integrated in terms of eMPLs,
discussing the details of the manipulations required to bring the integrals to the correct form.

The paper is organised as follows. 
We begin in Section~\ref{sec:eMPLs} with a review of eMPLs and their properties, in particular how to assign them
a concept of (uniform) transcendental weight,
in view of their usage in the next sections. 
We then move to explicit applications. In Section~\ref{sec:ttbar} we consider a family of 
two-loop non-planar three-point Feynman 
integrals, whose calculation is relevant
for $t\bar{t}$ and $\gamma \gamma$ production at the LHC. This family 
contains two elliptic master integrals, which we express in terms of eMPLs
by direct integration over the Feynman parameters. In Section~\ref{sec:ewff} 
we  consider a similar family of two-loop three-point functions,
 relevant for the computation of the electroweak form factor. The latter contains three 
 elliptic Feynman integrals, which we also explicitly integrate in terms
 of eMPLs. In Section~\ref{sec:kite} we show that  the same ideas can be applied also for the two-loop kite integral 
 with different internal masses.
 Finally, in Section~\ref{sec:conclusions} we draw our conclusions.



\section{Review of Elliptic Polylogarithms}
\label{sec:eMPLs}

Our goal in this paper is to show explicitly how the notion of elliptic multiple polylogarithms (eMPLs) developed in refs.~\cite{BrownLevin,Broedel:2017kkb,Broedel:2018iwv,Broedel:2018qkq} 
can be put into action for a wide range of Feynman integrals 
known not to be expressible in terms of ordinary MPLs. Before presenting the inner workings of this framework in specific examples, in this section we review the necessary concepts in the context of both ordinary and elliptic MPLs, and in particular the eMPLs introduced recently in ref.~\cite{Broedel:2018qkq}. The literature on eMPLs is vast, so here we content ourselves with summarising only the most important aspects necessary for the present calculations and refer the interested reader to refs.~\cite{Broedel:2017kkb,Broedel:2018qkq} for more detailed discussions.

Multiple polylogarithms are multi-valued functions defined recursively as iterated integrals over kernels which are rational functions with at most simple poles. 
The most well-known examples are the classic polylogarithms $\text{Li}_n (x)$, of which the logarithm is a special case,
\begin{align}
\bsp
\text{Li}_1(x)\,=\,-\log(1-x)\,,\quad \text{Li}_n(x)\,=\,\int_0^x\! \frac{dx^\prime}{x^\prime}\text{Li}_{n-1}(x^\prime)\ .
\esp
\end{align}
General MPLs are functions of many variables $a_i$ denoting the poles of the rational integration kernels, as well as the endpoint of the integration contour,
\beq
\label{eq:Mult_PolyLog_def}
G(a_1,\ldots,a_n;x)=\,\int_0^x\,\frac{d t}{t-a_1}\,G(a_2,\ldots,a_n;t)\,,\quad G(;x) = 1\ .
\eeq
They satisfy properties such as homotopy invariance (they do not depend on the details of the integration path and as such are functions only of its endpoint $x$) and shuffle relations,
\beq\label{eq:G_shuffle}
G(a_1,\ldots,a_k;x)\,G(a_{k+1},\ldots,a_{k+l};x) = \sum_{\sigma\in \Sigma(k,l)}G(a_{\sigma(1)},\ldots,a_{\sigma(k+l)};x)\,,
\eeq
where $\Sigma(k,l)$ stands for all order-preserving permutations of $\{a_1,\dots,a_k\} \cup \{a_{k+1},\dots,a_{k+l}\}$, called shuffles.

It is possible to assign notions of \textsl{length} and \textsl{weight} to MPLs. The length of an iterated integral (polylogarithmic or not)
 is always defined as the number of integrations, thus the length of an MPL 
 $G(a_1,\dots,a_n;x)$ is $n$. The notion of weight, however, is more subtle. 
 For MPLs, the weight is the same as the length, but as will become clear 
 once we discuss its elliptic version, this is not the general case. One can also assign a notion 
 of weight for constants which correspond to MPLs evaluated at special arguments. 
 While a constant has length zero (there are no integrals left to perform; see ref.~\cite{Broedel:2018qkq} for a detailed discussion), 
 the weight remembers that of the iterated integral it originated from. For example,
\begin{align}
\bsp
\log(-1) \,=\, i \pi \quad &\rightarrow \quad \text{Weight}(i\pi) = 1\,,\quad \text{Length}(i\pi) = 0\ ,\\
\zeta_n \,=\, \text{Li}_n(1)\quad &\rightarrow \quad \text{Weight}(\zeta_n) = n\,,\quad \text{Length}(\zeta_n) = 0\ .
\esp
\end{align}
Upon total differentiation, MPLs undergo a length drop, and their differential takes a particularly simple form,
\beq\bsp\label{eq:MPL_tot_diff}
d G(a_1,\ldots,a_n;z)&\, = \sum_{i=1}^nG(a_1,\ldots,\hat a_i,\ldots,a_n;z)\,d\ln{a_{i-1}-a_i\over a_{i+1}-a_i}\ ,
\esp\eeq
where we defined $a_0 \equiv 0 $ and $a_{n+1} \equiv z$. 
Functions whose total differential does not contain any homogeneous term are referred to as \textsl{unipotent}, and this concept will become important in the following discussions.

Elliptic generalisations of MPLs are functions which behave like MPLs but accommodate (in addition to the kernels $1/(x-a)$) functions which are rational in the variables $x$ and $y$ which define an elliptic curve, i.e.~$[x,y,1]\in \mathbb{CP}^2$ where $x$ and $y$ satisfy a polynomial equation $y^2 = P_n(x)$ of degree $n= 3,4$. For our purposes, we consider only the case with $n=4$ since the $n=3$ case can be seen a gauge-fixed version of the former and the examples we consider arise naturally as square roots of degree-four polynomials. Therefore, we are interested in iterated integrals of rational functions in the variables $(x,y)$ subject to the constraint
\beq
\label{eq:curve}
y^2 \,=\,P_4(x)\,=\, (x-a_1)(x-a_2)(x-a_3)(x-a_4)\ .
\eeq
The elements of the vector $\vec{a}\equiv(a_1,a_2,a_3,a_4)$ are referred to as the branch points of the elliptic curve.
The periods and quasi-periods of the elliptic curve are chosen according to
\begin{align}
\bsp
\label{eq:periods4}
\omega_1&\,=\, 2\,c_{4}\int_{a_2}^{a_3}\frac{dx}{y} = 2\,\textrm{K}(\lambda) \,, \\
\omega_2 &\,=\, 2\,c_{4}\int_{a_1}^{a_2}\frac{dx}{y} = 2i\,\textrm{K}(1-\lambda)\,,
\esp
\\
\bsp\label{eq:quasi-periods4}
\eta_1 &\,= -\frac{1}{2}\int_{a_2}^{a_3}dx\,\widetilde{\Phi}_4(x,\vec a) = \textrm{E}(\lambda) -\frac{2-\lambda}{3}\,\textrm{K}(\lambda)\,,\\
\eta_2 &\,= -\frac{1}{2}\int_{a_1}^{a_2}dx\,\widetilde{\Phi}_4(x,\vec a)= -i\,\textrm{E}(1-\lambda) +i\,\frac{1+\lambda}{3}\,\textrm{K}(1-\lambda)\ ,
\esp
\end{align}
where
\beq\label{eq:lambda4}
\lambda = \frac{a_{14}\,a_{23}}{a_{13}\,a_{24}} \ ,\qquad
c_{4} = \frac{1}{2}\sqrt{a_{13}a_{24}}\,,\qquad a_{ij} = a_i-a_j\,,
\eeq
and K and E denote the complete elliptic integrals of the first and second kind, respectively,
\beq
\textrm{K}(\lambda) = \int_0^1\frac{dt}{\sqrt{(1-t^2)(1-\lambda t^2)}}\,,\quad \textrm{E}(\lambda) = \int_0^1dt\,\sqrt{\frac{1-\lambda t^2}{1-t^2}}\ .
\eeq 
The function $\widetilde{\Phi}_4(x,\vec a)$ entering the integrand of the quasi-periods is defined as
\beq
\label{eq:tilde_Phi_4_def}
\widetilde{\Phi}_4(x,\vec a) \equiv \frac{1}{c_4\,y} \left( x^2 - \frac{s_1}{2}\,x + \frac{s_2}{6} \right)\,,
\eeq
where $s_n \equiv s_n(\vec a)$ denotes the $n^{\rm th}$ elementary symmetric polynomial in the branch points. 
The periods and quasi-periods are not independent and satisfy the Legendre relation,
\beq\label{eq:Legendre}
\omega_1\, \eta_2-\omega_2\,\eta_1\, =\, -i\pi\,.
\eeq

Since an elliptic curve $y$ is given in terms of a square root, it is important to make a choice for the signs of the branches of the square root $y$ which is consistent with the conventions for the periods and quasi-periods in eqs.~\eqref{eq:periods4} and \eqref{eq:quasi-periods4}.  In particular, we find it convenient to order the branch points such that $\omega_1 \in \mathbb{R}$ and $\omega_2 \in i \mathbb{R}$, whenever possible. Note that this implies that $\lambda$ defined in eq.~\eqref{eq:lambda4} lies between 0 and 1.
Moreover, in order to correctly define eMPLs we need to provide a prescription for how to perform the integrals in the regions of interest, namely on the real line and between branch points.
In this paper, we deal with Feynman-parameter integrals whose integrands are ratios of polynomials defined over the real numbers. Therefore, depending on the kinematic regime we wish to consider, the branch points can be real or complex so long as the polynomial $y^2 = P_4(x)$ is real. 
There are only three possible configurations for the branch points such that $y^2$ is real, which we consider in turn below (see fig.~\ref{fig:roots}):

\begin{figure}[h]
\centering
\includegraphics[width=\linewidth]{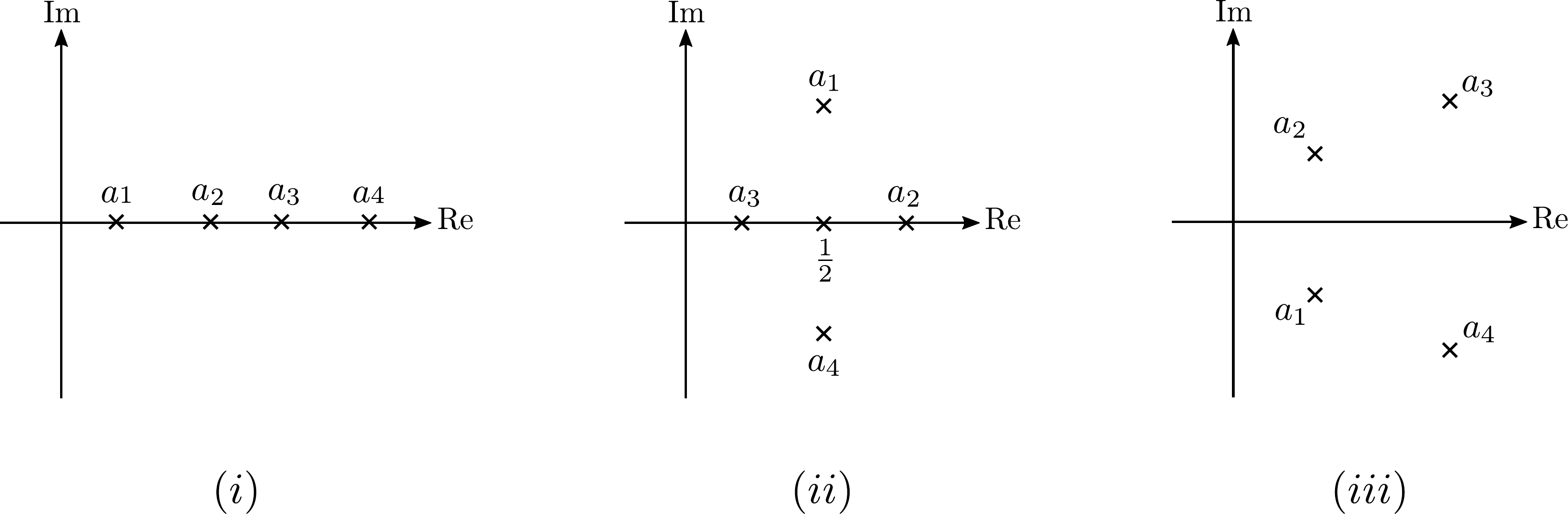}
\caption{Three possible configurations of the branch points in the complex plane such that $y^2 = P_4(x) \in \mathbb{R}$. $(i)$ All branch points are real and ordered. $(ii)$ Two branch points are real and two branch points are complex conjugate to each other. $(iii)$ All branch points are complex and pairwise complex conjugate to each other.}
\label{fig:roots}
\end{figure}

\begin{enumerate}

\item[$(i)$]  \textbf{All branch points are real.}\\
In this situation, we only need to consider integrations over the real axis. We order the branch points according to $a_1<a_2<a_3<a_4$  and fix the signs of the branches of the square root as
\beq\bsp
y\,=\,\sqrt{P_4(x)}&\,\equiv
\sqrt{|P_4(x)|}\times\left\{\begin{array}{ll}
-1\,,& x\le a_1\textrm{ or }x > a_4\,,\\
-i\,,& a_1<x\le a_2\,,\\
\phantom{-}1\,,& a_2<x\le a_3\,,\\
\phantom{-}i\,,& a_3<x\le a_4\,.
\end{array}\right.
\esp \label{eq:rsigns}
\eeq

\item[$(ii)$]  \textbf{Two branch points are real and two are complex conjugate to each other.}\\
The configuration of the branch points that feature in our applications is such that one can always impose the following ordering $\vec{a}=(a_1, a_2, a_3, a_4)$ where
\beq\bsp \label{eq:two_complex_roots}
a_1 \,=\, a_4^*  \in \mathbb{C} \,,\quad & \text{Im}(a_1) >0\, ,\\
a_2,\, a_3  \in \mathbb{R}\,,  \quad &  a_2  > a_3\,, \\
\text{Im}(a_1) \, > & \, \frac{a_2-a_3}{2}\ ,
\esp\eeq
where the branch points satisfy
\beq\bsp
\label{eq:relationstwocomplex}
a_4 = 1- a_1\,,\quad  a_3 \,=\,1-a_2\,,\\
\quad \text{Re}(a_1) \,=\, \text{Re}(a_4) \,=\, 1/2\ .
\esp\eeq
This choice seems particularly ad hoc, but we can motivate it as follows. In this situation, the polynomial $P_4(x)$ is negative on the real axis for $a_3 < x < a_2$ and for $x = 1/2 + i\, t,\, \text{Im}(a_4)<t<\text{Im}(a_1)$. We recall here that, in order to define the periods we need a prescription
for how to compute integrals between different branch points. 
In this case, in addition to a prescription to integrate along the real axis, it is enough to
supplement it with a prescription on the line $x = 1/2 + i\, t$ for ${\rm Im} (a_4) \leq t \leq {\rm Im} (a_1)$, see fig.~\ref{fig:roots}$(ii)$.
With this in mind, we define the elliptic curve in all regions of interest as
\beq
y = -i \sqrt{-(x-a_1)(x-a_2)(x-a_3)(x-a_4)}\ .
\eeq

\item[$(iii)$]  \textbf{All branch points are complex and pairwise complex conjugate.}\\
In this case, we order the roots such that
\beq
\bsp
a_1 \,=\, a_2^*\,,\quad a_3 \,=\, a_4^*\,,& \\
\text{Re}(a_1) \,<\,\text{Re}(a_3)\,,\quad \text{Im}(a_2),\,\text{Im}(a_3)\,>\,0\,,& \quad  \text{Im}(a_1),\,\text{Im}(a_4)\,<\,0\, .
\esp
\eeq
Since there is no branch cut on the real axis and $P_4(x)\geq0$ for $x \in \mathbb{R}$, 
we simply choose \cite{Broedel:2018qkq}
\beq
y\,=\, \sqrt{P_4(x)}\  .
\eeq
\end{enumerate}

eMPLs were originally defined in refs.~\cite{BrownLevin,MatthesThesis,Broedel:2014vla} as iterated integrals on a complex torus. This description is related to the one presented here through the relation between elliptic curves defined by the equation $y^2 = P_4(x)$ and a torus defined as the complex plane quotiented by a two-dimmensional lattice 
$\Lambda = \mathbb{Z}\, \omega_1 + \mathbb{Z}\, \omega_2$. The ratio of the two lattice periods $\tau = \omega_2/\omega_1$ is called the modular parameter and it is easy to see that the lattice $\Lambda$ remains invariant under modular transformations $SL(2,\mathbb{Z})$ mixing the two periods. Modular transformations act on $\tau$ as M\"{o}bius transformations. The way to map an elliptic curve to its equivalent torus description is via the function~\cite{Broedel:2017kkb},
\beq
\kappa(z)\,=\,\frac{-3 a_1 a_{13}a_{24}\wp(z)+a_1^2 \bar{s}_1-2 a_1 \bar{s}_2+ 3\bar{s}_3}{-3 a_{13}a_{24}\wp(z)+a_1^2 -2 a_1 \bar{s}_1+ 3\bar{s}_2}\, .
\eeq
Here $z$ is a variable on the torus, $\wp(z)$ is the Weierstrass $\wp$-function, $\bar{s}_n\equiv s_n(a_2, a_3, a_4)$ and the $s_n$ are the symmetric polynomials defined below eq.~\eqref{eq:tilde_Phi_4_def}.
The $\kappa$-function satisfies a differential equation which is identical to the definition of the elliptic curve in eq.~\eqref{eq:curve}, namely
\begin{align}
(c_4 \kappa^\prime(z))^2 = P_4(\kappa(z))\ ,
\end{align}
and thus one may identify $(x,y) \leftrightarrow (\kappa(z), c_4 \kappa^\prime(z))$. The inverse of the $\kappa$-function is known as Abel's map, which takes a point $(x, y)$ on the elliptic curve to a point $z_x$ on the complex torus,
\beq\label{eq:Abel}
z_x = \frac{c_4}{\omega_1}\int_{a_1}^x\frac{dx^\prime}{y} = \frac{\sqrt{a_{13}a_{24}}}{4\,\textrm{K}(\lambda)}\int_{a_1}^x\frac{dx^\prime}{y}\,.
\eeq

Since elliptic curves are isomorphic to complex tori, eMPLs can be described as iterated integrals over functions related to the torus, and were originally defined as such in refs.~\cite{BrownLevin,MatthesThesis,Broedel:2014vla}. In this context, eMPLs are defined as iterated integrals given by
\beq\label{eq:gamt_def}
\gamtt{n_1 &\ldots& n_k}{z_1 & \ldots & z_k}{z}{\tau} = \int_0^zdz'\,g^{(n_1)}(z'-z_1,\tau)\,\gamtt{n_2 &\ldots& n_k}{z_2 & \ldots & z_k}{z'}{\tau}\,,
\eeq
where the integration kernels are the coefficients in the expansion of the \textsl{Kronecker-Eisenstein} series $F(z,\alpha,\tau)$,
\beq\label{eq:Eisenstein-Kronecker}
F(z,\alpha,\tau) = \frac{1}{\alpha}\,\sum_{n\ge0}g^{(n)}(z,\tau)\,\alpha^n = \frac{\theta'_1(0,\tau)\,\theta_1(z+\alpha,\tau)}{\theta_1(z,\tau)\,\theta_1(\alpha,\tau)}\,,
\eeq
and $\theta_1(z,\tau)$ is the odd Jabobi theta function with $\theta_1^\prime(z,\tau)$ denoting a derivative with respect to its first argument. 

The eMPLs \eqref{eq:gamt_def} behave similarly to ordinary MPLs in that they also form a shuffle algebra and are unipotent. Moreover, they are pure according to the definition of ref. \cite{Broedel:2018qkq}, namely: \textsl{A function is called pure if it is unipotent and its total differential involves only pure functions and one-forms with at most logarithmic singularities}. 

In the calculation of Feynman integrals that evaluate to functions of the elliptic kind, 
the representation of elliptic polylogarithms in terms of 
a polynomial equation $y^2 = P(x)$ appears more naturally than the torus picture. 
Therefore in this paper we use the definition of pure eMPLs on the elliptic curve recently put forward in ref.~\cite{Broedel:2018qkq}.
They are defined as iterated integrals of kernels that are rational functions on the elliptic curve with at most logarithmic singularities in all variables,
\beq
\label{eq:cE4_def}
\cEfe{n_1 & \ldots & n_k}{c_1 & \ldots& c_k}{x}{\vec{a}} = \int_0^xdt\,\Psi_{n_1}(c_1,t,\vec a)\,\cEfe{n_2 & \ldots & n_k}{c_2 & \ldots& c_k}{t}{\vec a}\, .
\eeq
In contrast with MPLs, for the elliptic case the requirement that all integrations over rational functions on the elliptic curve close on the same space of functions put together with the requirement that all integrals must have at most logarithmic singularities leads to an infinite tower of independent kernels $\Psi_n$ for $n\in\mathbb{Z}$. This fact can also be seen from the torus description, where an infinite number of kernels are generated by eq.~\eqref{eq:Eisenstein-Kronecker}. In particular, the kernels in eq.~\eqref{eq:cE4_def} depend on a certain kind of functions which are themselves transcendental, namely
\beq
\label{eq:Z4_def}
Z_4(x,\vec a) \equiv \int_{a_1}^xdx'\,\Phi_4(x',\vec a) \,,\qquad \textrm{with} \qquad \Phi_4(x,\vec a)\equiv\widetilde\Phi_4(x,\vec a) +4c_{4}\, \frac{\eta_1}{\omega_1}\,\frac{1}{y}\,,
\eeq
and $\widetilde\Phi_4(x,\vec a) $ given in eq.~\eqref{eq:tilde_Phi_4_def}. 

The kernels $\Psi_n$ entering the eMPLs in eq.~\eqref{eq:cE4_def} are spelled out below for $|n| = 0,1,2$. 
Higher values of $n$ do not appear in the present applications since the corresponding functions 
would satisfy higher-order differential equations. Before writing down the expressions for the kernels, 
we introduce some functions which appear as ingredients. The first is the function $Z_4(x,\vec a)$ 
defined in eq.~\eqref{eq:Z4_def}. Likewise, an important element is the image of the point $x=-\infty$ under Abel's map \eqref{eq:Abel},
\beq
z_\ast = \frac{c_4}{\omega_1}\int_{a_1}^{-\infty}\frac{dx^\prime}{y} \ .
\eeq
It is possible to represent $z_\ast$ in terms of elliptic integrals. In the situation where all roots are real and ordered, it is given by \cite{Broedel:2018qkq},
\beq\label{eq:zstar}
z_{\ast} = \cZ_{\ast}(\alpha,\lambda)\equiv  \frac{1}{2} - \frac{\textrm{F}(\sqrt{\alpha} | \lambda)}{ 2\, \textrm{K}(\lambda)}\,, \qquad \alpha = \frac{a_{13}}{a_{14}}\,,
\eeq
and for other configuration of the branch points (or complex ones), $z_\ast$ may pick up a minus sign depending on the conventions for the branches of the square root.
Finally, the kernels entering the pure eMPLs depend on the function $G_{\ast}(\vec a)$, which is simply 
the image of the point $z_\ast$ under $g^{(1)}$ (see eq.~\eqref{eq:Eisenstein-Kronecker}),
\beq
\label{eq:G_infty_def}
G_{\ast}(\vec a) \equiv \frac{1}{\omega_1}\,g^{(1)}(z_{\ast},\tau)\,.
\eeq
As shown in ref.~\cite{Broedel:2018qkq}, $G_{\ast}$ can be integrated explicitly 
in terms of (incomplete) elliptic integrals of the first and second kind. 
In the situation where the branch points $\vec{a}$ are real and ordered 
according to $a_1 < a_2 < a_3 < a_4 $ one finds
\beq\bsp\label{eq:G_infty_KE}
G_{\ast}(\vec a)
&\, = \left(\frac{2  \eta_1}{ \omega_1}-\frac{\lambda }{3}+\frac{2}{3}\right) \textrm{F}\!\left(\sqrt{\alpha}|\lambda \right)-\textrm{E}\!\left(\sqrt{\alpha }|\lambda \right)+\sqrt{\frac{\alpha  (\alpha  \lambda -1)}{\alpha -1}}\,.
\esp\eeq
In the special case where the point $z_\ast$ is of the 
form\footnote{The situation with $a,b\in\mathbb{Q}$ is common in applications, 
and a point on the elliptic curve of this form is called a \textsl{torsion point}.
} 
\beq
z_\ast \,=\, a + b \,\tau(\lambda) \ , \label{eq:zstarrat}
\eeq
for $a$ and $b$ constants, then $G_\ast(\vec{a})$ admits an even simpler form, namely
\begin{align}
G_\ast(\vec{a})&\,=\frac{(1-\lambda)\left[\lambda\,  \alpha '(\lambda ) 
+ \alpha\right]}{\sqrt{ \alpha (1-\alpha )  (1-\alpha  \lambda )}}
-b\,  \frac{ 2\pi i }{ \omega_1}\ . \label{eq:Gstarrat}
\end{align}
This follows because eq.~\eqref{eq:zstar} together with eq.~\eqref{eq:zstarrat} imply that
$\alpha = \alpha(\lambda)$.

At last, we are now ready to write down the expressions for the kernels. 
For $n=0$, there is only one kernel,
\beq\label{eq:pure_psi0}
\Psi_0(0,x,\vec a) \,=\, \frac{c_4}{\omega_1\,y}\,.
\eeq
For $n=1$, we have instead four kernels  (with $c\neq \infty$)
\begin{align}
\nonumber \Psi_1(c,x,\vec a) &\,=\, \frac{1}{x-c}\,, \\
\label{eq:pure_psi1}\Psi_{-1}(c,x,\vec a) &\,= \, \frac{y_c}{y(x-c)} + Z_4(c,\vec a)\,\frac{c_4}{y}\,,\\
\nonumber\Psi_{1}(\infty,x,\vec a) &\,= \, -Z_4(x,\vec a)\,\frac{c_4}{y}\,,\\
\nonumber\Psi_{-1}(\infty,x,\vec a) &\,= \, \frac{x}{y}   -\frac{1}{y} \left[{a_1}+ 2c_4\, G_{\ast}(\vec a)\right]\,,
\end{align}
where $y_c \equiv \sqrt{P_4(c)}$.
Finally, for $n=2$, we have (with $c\neq \infty$), 
\begin{align}
\bsp \label{eq:pure_psi2}
\Psi_{2}&(c,x,\vec{a})\,=\, \frac{\omega _1}{12 (x-c)}\Big[ \frac{6 \left(a_1-c\right)Z_4(x,\vec{a})}{x-a_1} + \frac{6 y_c Z_4(c,\vec{a})\left(-a_1+2 c-x\right)}{(c-a_1) y}\\
&- \frac{2 (c-x) (y_c (3 a_{13} a_{24} (Z^{(2)}_4(c,\vec{a})+Z^{(2)}_4(x,\vec{a}))+a_1 (2 (a_2+ a_3+ a_4)-3 (c+x))}{c_4 y_c y} \\
& \frac{-a_2 a_3-(a_2+a_3) a_4+3 c x)+3 \sqrt{a_{13} a_{24}} (c-a_2) \big(c-a_3) (c-a_4) Z_4(c,\vec{a})\big)}{c_4 y_c y} \Big]\ ,\\
\Psi_{-2}&(c,x,\vec{a})\,=\,  \frac{\omega _1}{2 (x-c)} \Big[\frac{- y_c Z_4(x,\vec{a})}{y}+Z_4(c,\vec{a})\left(\frac{c_4 (c-x)Z_4(x,\vec{a})}{y}+1\right)\Big]\ ,\\
\Psi_{2}&(\infty,x,\vec a) \,= \, \frac{1}{4} \frac{\omega _1}{c_4 y} \Big[2 a_{13} a_{24} Z_4^{(2)}(x,\vec{a})+4 c_4 G_\ast(\vec{a}) \left(a_1-x\right)+a_{13} a_{24} G_\ast(\vec{a})^2\\
&+2 (-a_1+a_2+a_3+a_4) x+a_1^2-a_2 a_3-a_2 a_4-a_3 a_4-2 x^2\Big]-\frac{1}{2} \omega _1\frac{Z_4(x,\vec{a})}{a_1-x}\ ,\\
\Psi_{-2}&(\infty,x,\vec a) \,= \, \frac{\omega _1}{2 c_4}\Big[1+\frac{Z_4(x,\vec{a})\left(2 c_4 \left(a_1-x\right)+a_{13} a_{24} G_\ast(\vec{a})\right)}{2 y}\Big]\ ,
\esp
\end{align}
where $Z_4^{(2)}(x,\vec{a})$ stands for a degree-two polynomial in $Z_4(x,\vec{a})$,
\beq\bsp
&Z_4^{(2)}(x,\vec{a})\,=\,\frac{1}{8} Z_4(x,\vec{a})^2+\frac{\left(a_2-x\right) \left(a_3-x\right) \left(a_4-x\right) Z_4(x,\vec{a})}{c_4 x}\\
&+\frac{-3 a_3 x-3 a_4 x-a_1 \left(a_2+a_3+a_4-3 x\right)+a_2 \left(2 a_3+2 a_4-3 x\right)+2 a_3 a_4+3 x^2}{6 a_{13} a_{24}}\ .
\esp\eeq

We conclude this short exposition of pure eMPLs with a comment: much like with ordinary MPLs, one can associate a concept of length and of weight to eMPLs and to quantities which arise from 
evaluating eMPLs at special points, for example the periods and quasi-periods of the elliptic curve defined in 
eqs.~\eqref{eq:periods4} and \eqref{eq:quasi-periods4}.  In summary we have the values shown in Table~\ref{tab:lw-periods}.

\begin{table}[h]
\centering
\begin{tabular}{c|c|c}
 & Length & Weight \\[3pt]\hline & & \\[-5pt]
 $\gamtt{n_1 &\ldots& n_k}{z_1 & \ldots & z_k}{z}{\tau}$ & $k$ &   $\sum_{i=1}^k n_i$ \\[5pt]
$ \cEfe{n_1 & \ldots & n_k}{c_1 & \ldots& c_k}{x}{\vec{a}}$ & $k$ &  $\sum_{i=1}^k |n_i| $\\[5pt]
$\omega_1$, $\eta_1$ & 0 & 1 \\[5pt]
$\tau$ & 1 & 0 
\end{tabular}
\caption{Length and weight of eMPLs and related constants.}
\label{tab:lw-periods}
\end{table}

Using the formalism revised in this section, in the rest of this paper we will show how certain 
Feynman integrals which evaluate to functions beyond 
MPLs can be brought to neat expressions in terms of combinations of pure 
eMPLs \eqref{eq:cE4_def} of uniform weight by direct integration of their Feynman parametrisation.


\section{A non-planar triangle with a massive loop}
\label{sec:ttbar}

\begin{figure}[h]
	\centering
	\includegraphics[width=0.3\linewidth]{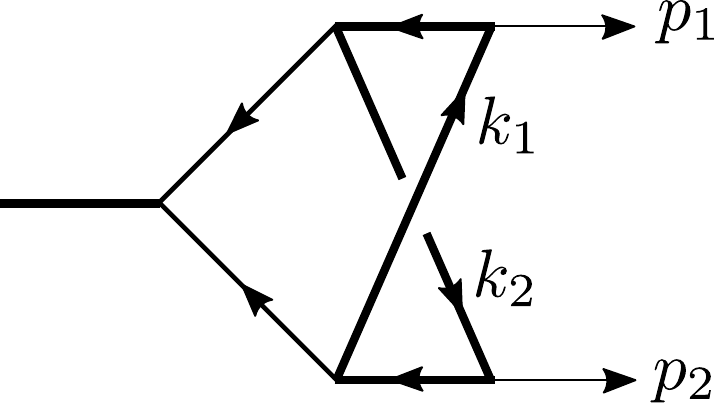}
	\caption{Triangle with massive loop.}
	\label{fig:trimassloop}
\end{figure}

In this first application, we consider the family of  two-loop non-planar three-point functions  
with a massive loop shown in fig.~\ref{fig:trimassloop}. These integrals involve massless propagators and a massive loop with four propagators with mass $m$. 
Two external legs are massless, i.e.~$p_1^2=p_2^2=0$ and we set our kinematic variable 
$q^2= (p_1 + p_2)^2$. The family of two-loop integrals is then
\begin{align}
\label{eq:trimass}
I_{a_1,\dots,a_7}\,=\,-\frac{1}{\pi^D}\int\! d^D k_1 d^D k_2 \frac{D_7^{-a_7}}{\prod_{i=1}^6 D_i^{a_i}}\ ,
\end{align}
where the $a_i$ are the exponents of the propagators, and we consider only integrals with 
$a_7<0$. The propagators are
\begin{equation}
\begin{split}
D_1\,&=\, k_1^2-m^2, \quad D_3\,=\,(k_1-p_1)^2-m^2,\quad D_5\,=\,(k_1-k_2-p_1)^2 , \\
 D_2\,&=\,k_2^2-m^2, \quad  D_4\,=\, (k_2-p_2)^2-m^2,\quad D_6\,=\,(k_2-k_1-p_2)^2,\quad D_7 =k_1 \cdot p_2 .
\end{split}
\end{equation}

The family of integrals in eq.~\eqref{eq:trimass} was studied in ref.~\cite{vonManteuffel:2017hms} by means of the differential-equation method, where it was shown that there are two master integrals for the top topology which satisfy a coupled two-dimensional system.\footnote{For a numerical implementation of this integral, see ref.~\cite{Bonciani:2018uvv}.} We choose as basis integrals
\beq
\label{eq:mastersttbar}
M_1\,=\,I_{1,1,1,1,1,1,0}\,,\quad M_2 \,=\, I_{2,1,1,1,1,1,0}\, ,
\eeq 
where we note that both integrals are finite in  $D = 4$ space-time dimensions.
In ref.~\cite{vonManteuffel:2017hms}, the solution of the system was presented as an expansion in the dimensional-regularisation parameter $\epsilon = (4-D)/2$ with coefficients being integrals over elliptic integrals of the first and second kind multiplied by ordinary MPLs and rational functions.

In this paper, we apply the framework of eMPLs 
on the elliptic curve reviewed in Section~\ref{sec:eMPLs} in order to obtain 
analytic expressions for these Feynman integrals
and study their properties. In particular, we show that once the two master integrals 
in the top sector are expressed in terms
of our class of functions, one can easily transform them to a new basis of 
master integrals which are explicitly pure, similarly to what one would do if they were
standard MPLs. 
To this end, we approach the problem through direct integration 
over the Feynman parameter representation of the integrals and show that all 
integrations up to the penultimate can be performed in terms of ordinary 
MPLs with algebraic arguments that can in general depend on an elliptic curve. 
In order to perform the last integration, we employ partial fractioning to rewrite 
the integrand as kernels of the eMPLs defined in eq.~\eqref{eq:cE4_def} (see eqs.~\eqref{eq:pure_psi0}, \eqref{eq:pure_psi1} and \eqref{eq:pure_psi2}).
We now look in detail at the computation of the two master integrals.

\subsection{First master integral}
\label{sec:ttbar-first-master}

We start by considering the first master integral $M_1$, defined in eq.~\eqref{eq:mastersttbar}.
Our task is to find an order of integration of the Feynman parameters such that linear 
reducibility is achieved for all integrations except the last one, which will in turn require the introduction of eMPLs.
The Feynman parameter representation of the first master integral is given by
\begin{align}
\label{eq:3pt-integral-ml}
M_1\,=\,  \Gamma(6-D)  \, \int_{0}^\infty \prod_{i=1}^6 d  x_i  \, \delta(1-\sum_{i=1}^6 x_i) \,\frac{\mathcal{U}^{6-3D/2}}{\mathcal{F}^{6-D}}\ ,
\end{align}
where $\mathcal{U}$ and $\mathcal{F}$ are the Symanzik polynomials associated with the graph. 
Since this integral is finite, it can be evaluated directly in $D=4$, where the $\mathcal{U}$-polynomial drops out. 

The next step is to apply the Cheng-Wu theorem \cite{Cheng:1987ga} in order to find a particular 
parametrisation of the integral such that as many Feynman parameter integrations as possible can be 
done in terms of ordinary MPLs.
In practice, the Cheng-Wu theorem allows one to exchange the integration 
domain $\Delta$ of a Feynman integral by using any subset of propagators $\Sigma$ such that
\begin{equation}
\Delta_\Sigma = \Big\{ x_i > 0 \Big| \sum_{i\in\Sigma}x_i = 1 \Big\} \ .
\end{equation}
Here we choose to apply the theorem with $\Sigma = \{1,2,3,4\}$. 
This particular ordering was used first in ref.~\cite{Hidding:2017jkk} to write a one-fold integral representation
for a similar integral.\\[5pt]
In doing so, the integral \eqref{eq:trimass} becomes
\begin{align}
M_1\,=\,2\, \int_{0}^\infty \! dx_6 \int_{0}^\infty \! dx_5 \, \int_{0}^{1} \! dx_4 \int_{0}^{1-x_4} \! dx_2   \int_{0}^{1-x_2-x_4} \! dx_3 \int_{0}^{1-x_2-x_3-x_4} \! dx_1 \,\frac{\delta(1-\sum_{i=1}^4x_i)}{\mathcal{F}^{2}}\ ,
\end{align}
where the polynomial $\mathcal{F}$ with $x_1 = 1-x_2-x_3-x_4$ reads
\begin{align}
\begin{split}
\mathcal{F}\,=\,& (-q^2) \Big[ a \left(x_2^2+\left(2 x_4-1\right) x_2+x_4^2-x_4-x_5-x_6\right)+x_5 x_4^2\\
&+\left(x_2+x_3-1\right) x_5 x_4-\left(x_2 x_3+x_5\right) x_6\Big] \ ,
\end{split}
\end{align}
and we have factored out $(-q^2)$ and encoded the kinematic dependence in the dimensionless variable 
\begin{equation}
\label{eq:adef}
a\,\equiv\,\frac{m^2}{(-q^2)}\ .
\end{equation} 
From now on we set $m^2=1$ for simplicity and restore its dependence at the end using dimensional analysis.
The integrals over $x_6,\,x_5$ and $x_3$ can be done easily in terms of MPLs~\eqref{eq:Mult_PolyLog_def} as the integrand is linearly reducible in these variables. This results in a two-fold integral,
\begin{align}
\label{eq:M1-twofold}
\begin{split}
M_1\,= \,2 a^2 \, &\int_{0}^{1} dx_4 \int_{0}^{1-x_4} dx_2 \frac{1}{a (x_2+x_4)+(x_2+x_4-1) x_2 x_4}  \Big\{G(1-x_4;x_2) \Big[G\Big(\frac{a}{x_4};1-x_2-x_4\Big)\\
&-G\Big(1-\frac{a}{x_2}-x_2-x_4;1-x_2-x_4\Big)\Big]-G\Big(1+\frac{a}{x_4}-x_4;x_2\Big) G\Big(\frac{a}{x_4};1-x_2-x_4\Big)\\
&+G\Big(1-\frac{a}{x_2}-x_2-x_4;1-x_2-x_4\Big) \Big[G\Big(1+\frac{a}{x_4}-x_4;x_2\Big)+G((x_4-1) x_4;a)\\&
-G(-x_4;x_2)\Big] -G((x_4-1) x_4;a) G\Big(\frac{a}{x_4};1-x_2-x_4\Big) +G(-x_4;x_2) G\Big(\frac{a}{x_4};1-x_2-x_4\Big)\\&
+G\Big(1-\frac{a}{x_2}-x_2-x_4,1+\frac{a}{x_4}-x_2-x_4;1-x_2-x_4\Big)\\&-G\Big(\frac{a}{x_4},-\frac{a}{x_2};1-x_2-x_4\Big)
+G\Big(1-\frac{a}{x_2}-x_2-x_4,-\frac{a}{x_2};1-x_2-x_4\Big)\\&
-G\Big(\frac{a}{x_4},\frac{a}{x_4}-x_2-x_4+1;1-x_2-x_4\Big)\Big\}\ .
\end{split}
\end{align}
The next integration to be done is over $x_2$. We notice immediately that the
overall rational pre-factor is quadratic both in $x_2$ and $x_4$. Indeed, as we will see
below, upon integration in either variable this will give rise to a square root of a polynomial
of degree four in the other one, defining an elliptic curve.
On top of this, performing a study of the symbol alphabet of the combination of MPLs
in eq.~\eqref{eq:M1-twofold}, we find the following letters,
\begin{align}
\begin{split}
&\{a,\,x_2,\,x_4,\,x_2+x_4,\,1-x_2-x_4,\, a+x_2(1-x_2-x_4),\\
a-&x_2(1-x_2-x_4),\,a+x_4(1-x_2-x_4),\,a-x_4(1-x_2-x_4)\}\ .
\end{split}
\end{align}
The alphabet above involves quadratic letters both in $x_2$ and $x_4$, such that
by rewriting the MPLs in the form $G(\dots, x_2)$, one would in general be left with MPLs involving multiple square roots involving $x_4$.
The presence of these additional square roots
could prevent us from performing the last integration
in terms of eMPLs algorithmically. 
However, it is easy to realise that 
after a simple change of variables $x_2 \rightarrow \bar{x}_2 \equiv x_2 + x_4$
 all symbol letters become linear in $x_4$,
\begin{align}
\begin{split}
\{a,&\,x_4,\,1-\bar{x}_2,\,x_4-\bar{x}_2,\,\bar{x}_2,\,a+x_4(1-\bar{x}_2),\,a-x_4(1-\bar{x}_2)\\
&a+(1-\bar{x}_2)(\bar{x}_2-x_4),\,a-(1-\bar{x}_2)(\bar{x}_2-x_4)\}\ , \label{eq:symlin}
\end{split}
\end{align}
such that if we perform the change of variables and 
exchange the order of integration in eq.~\eqref{eq:M1-twofold} as follows,
\begin{align}
\int_{0}^{1} dx_4 \int_{0}^{1-x_4} dx_2 \,=\, \int_{0}^{1} dx_4 \int_{x_4}^{1} d\bar{x}_2 \,=\,\int_{0}^{1} d \bar{x}_2  \int_{0}^{\bar{x}_2} dx_4\ ,
\end{align}
we expect to be able to perform the integral in $x_4$ without introducing any additional
square roots in $\bar{x}_2$.
Note that, while this transformation linearises the overall symbol of the integrand, individual MPLs may still involve quadratic symbol letters that cancel out in the combination. 
Indeed, in this case it turns out that in order to rewrite the individual MPLs one needs
to introduce the square-root valued letters
\begin{align}
\label{eq:rs}
r_{\pm} = \frac{1}{2}(1-\sqrt{1 \pm  4a})\ .
\end{align}
These letters enter in identities of the type
\begin{align}
\begin{split}
G\left(\frac{a}{x_4}-x_4+1;\bar{x}_2-x_4\right)\,=\,&
G\left(-\frac{a}{1-\bar{x}_2};x_4\right)-G\left(\rmp;x_4\right)-G\left(\rpp;x_4\right)\\[5pt]
G\left(\left(x_4-1\right) x_4;a\right)\,=\,& G\left(\rmp;x_4\right)+G\left(\rpp;x_4\right)+\log\left(\frac{a}{x_4(1-x_4)}\right)\ .
\end{split}
\end{align}
Once these identities are inserted back into eq.~\eqref{eq:M1-twofold}, 
the dependence on $r_{\pm}$ cancels out,
as expected from eq.~\eqref{eq:symlin}.
Note however that, for this particular example, 
the cancellation of $r_{\pm}$ at this stage is not required for the integration 
algorithm to go through, since $r_{\pm}$ do not depend on the remaining integration
variable $\bar{x}_2$. 

By performing these manipulations and integrating over $x_4$ 
through the recursive definition of MPLs \eqref{eq:Mult_PolyLog_def}  we arrive at a 
one-fold integral in the variable $\bar{x}_2$ given by
\begin{align}
\label{eq:last_int}
\begin{split}
&M_1\,=\, \frac{2 a^2 }{3}\, \int_0^1 
\frac{d\bar{x}_2}{y} \\ &
\times \Bigg[ 
6 \Bigg(G\left(\left(\bar{x}_2-1\right) \bar{x}_2;a\right)
\left(G_-\left(-\frac{a}{\bar{x}_2-1};\bar{x}_2\right)+2
G_-\left(\bar{x}_2;\bar{x}_2\right)\right) \\& +G\left(0;\bar{x}_2\right) \left(2
G_-\left(\bar{x}_2;\bar{x}_2\right)-G_-\left(\frac{a}{\bar{x}_2-1}+\bar{x}_2;\bar{x}_2\right)\right)
-G\left(1;\bar{x}_2\right) G_-\left(\frac{a}{\bar{x}_2-1}+\bar{x}_2;\bar{x}_2\right) \\&
+2
G_-\left(0,\frac{a}{\bar{x}_2-1};\bar{x}_2\right)+G_-\left(-\frac{a}{\bar{x}_2-1},\frac{a-\bar{x}_2^2+\bar{x}_2}{1
	-\bar{x}_2};\bar{x}_2\right) \\&
+2
G_-\left(\bar{x}_2,\frac{a-\bar{x}_2^2+\bar{x}_2}{1-\bar{x}_2};\bar{x}_2\right)+G_- \left(\frac{a}{\bar{x}_2-1}+\bar{x}_2,\frac
{a}{\bar{x}_2-1};\bar{x}_2\right) \\
& -2 \log (a) G_-\left(\bar{x}_2;\bar{x}_2\right)+\log (a)
G_-\left(\frac{a}{\bar{x}_2-1}+\bar{x}_2;\bar{x}_2\right)+2 G\left(1;\bar{x}_2\right)
G_-\left(\bar{x}_2;\bar{x}_2\right)\Bigg)
\\& -G_-\left(\bar{x}_2\right) \Bigg( 6 G\left(0;\bar{x}_2\right)
G\left(\left(1-\bar{x}_2\right) \bar{x}_2;a\right)+6 G\left(1;\bar{x}_2\right)
G\left(\left(1-\bar{x}_2\right) \bar{x}_2;a\right) \\& 
+6 G\left(0,\left(1-\bar{x}_2\right) \bar{x}_2;a\right)+6
G\left(0,\left(\bar{x}_2-1\right) \bar{x}_2;a\right)-6 \log (a) G\left(\left(1-\bar{x}_2\right)
\bar{x}_2;a\right)+\pi ^2\Bigg) \Bigg]  \ ,
\end{split}
\end{align}
where $y$ is the square root of a quartic polynomial, as anticipated,
\beq
y^2 = P_4(\bar{x}_2) = \bar{x}_2 (\bar{x}_2-1)(\bar{x}_2-b_+)(\bar{x}_2-b_-)\ ,
\eeq
and as such defines an elliptic curve with branch points
\beq
\label{eq:rootsttbar}
\vec{b}\,=\,(b_-,\,1,\,0,\,b_+) \,,\quad b_\pm \,=\, \frac{1}{2}(1\pm\sqrt{1-16a})\ .
\eeq

The ordering of the branch points in $\vec{b}$ is chosen following the prescription in eq.~\eqref{eq:two_complex_roots} for the kinematic region where $a>1/16$, so that $b_- = (b_+)^*$ are complex. This choice is made for convenience, as it is simpler to perform a numerical evaluation of the final result when the branch points and potential poles are not on the real axis.
In eq.~\eqref{eq:last_int} we have also introduced the shorthand notation for symmetric and anti-symmetric combinations of MPLs depending on the elliptic curve,
\beq
G_{\pm}(\vec{n}, x) \equiv \frac{G(\Rp, \vec{n}, x) \pm G(\Rm, \vec{n}, x)}{2}\ ,
\eeq
with the variables $R_{\pm}$ given by
\begin{align}
\begin{split}
R_{\pm} &\,\equiv\, 
\frac{- \bar{x}_2(1-\bar{x}_2) \pm 
	\sqrt{P_4(\bar{x}_2)}
}{2 \left(\bar{x}_2-1\right)}\ .
\end{split}
\end{align}

In order to compute the integral in eq.~\eqref{eq:last_int}, our next task is to recast its integrand in terms of eMPLs  
whose dependence on the integration variable is of the form $\mathcal{E}_4(\dots;\bar{x}_2)$ multiplied by eMPL kernels (see eqs.~\eqref{eq:pure_psi0}, \eqref{eq:pure_psi1} and \eqref{eq:pure_psi2}). 
This way, the last integral in the variable $\bar{x}_2$ can be performed trivially using the recursive definition of eMPLs 
\eqref{eq:cE4_def}.
This is a bottom-up procedure in the length of the eMPLs: at a given length $L = n$, it
can be achieved through a sequence of four steps:
\begin{enumerate}
\item Differentiation in the variable $\bar{x}_2$, which in general 
produces rational functions on the elliptic curve\footnote{We recall that a rational function 
on the elliptic curve is defined as a rational function in two variables, $R(x,y)$, with 
the constraint $y = \sqrt{P_4(x)}$.}
and eMPLs of length $L = n-1$. Since we are working bottom-up, 
we can assume that all eMPLs of length $L=n-1$ are already known in the form $\mathcal{E}_4(\dots;\bar{x}_2)$.
\item Partial fractioning of the derivative to cast it as 
a linear combination of eMPL kernels, times eMPLs of length $L=n-1$.
\item Finding the primitive in $\bar{x}_2$ using the recursive definition of the eMPLs.
\item Fixing of the integration constant comparing the original expression and the 
new one for a fixed (possibly simple)
 value of $\bar{x}_2$.
\end{enumerate}
This procedure guarantees that all eMPLs are of the form $\mathcal{E}_4(\dots;\bar{x}_2)$ times an elementary eMPL kernel and thus can be trivially integrated using eq.~\eqref{eq:cE4_def}.
Note that this is a standard procedure for ordinary MPLs and 
the only difference here is that we need to identify and use the elliptic kernels instead.

To illustrate the mechanism, let us first consider the MPLs that do not depend on the elliptic curve, i.e.~$G(\vec{n};x)$ where the endpoint $x$ and the letters $\vec{n}$ are either constants or depend on the square-root letters defined in eq.~\eqref{eq:rs}, but not on $y$. In this situation, a derivative in $\bar{x}_2$ plus partial fractioning will lead to an integrand which depends only on polylogarithmic kernels of the form $\frac{1}{\bar{x}_2-c}$, with $c$ independent of $\bar{x}_2$. As such, it can simply be integrated back to an MPL of the form $G(\vec{m},\bar{x}_2)$ with $\vec{m}$ independent of $\bar{x}_2$.

In a similar fashion, for the situation where the MPLs depend on the elliptic curve, the procedure will lead to a 
combination of eMPL kernels which can then be integrated to $\mathcal{E}_4$ functions.  As an example, consider the weight-one function
\begin{align}
\Gm(\bar{x}_2) = \frac{G(\Rp, \bar{x}_2) - G(\Rm, \bar{x}_2)}{2}
= \frac{1}{2}\log\left( \frac{ \bar{x}_2(1-\bar{x}_2) + \sqrt{P_4(\bar{x}_2)}}{ \bar{x}_2(1-\bar{x}_2) - \sqrt{P_4( \bar{x}_2)}} \right)\ .
\end{align}
Taking a derivative with respect to $\bar{x}_2$ yields 
\begin{align}
\frac{\partial}{\partial \bar{x}_2} \Gm(\bar{x}_2)\,&=\,\frac{1 - 2 \bar{x}_2}
{2\sqrt{ \bar{x}_2\left(\bar{x}_2-1\right)(4 a-\bar{x}_2 (1 - \bar{x}_2 ))}} \,=\,  \frac{1}{2y}  - \frac{\bar{x}_2}
{y}\ .
\end{align}
The next step is to integrate the above expression back in terms of $\mathcal{E}_4$ functions. To this end, we rewrite the integrand in terms of the kernels defined in eqs.~\eqref{eq:pure_psi0} and \eqref{eq:pure_psi1},
\begin{align}
\begin{split}
\Gm(\bar{x}_2) \,&=\, \int_0^{\bar{x}_2} \frac{dt}{y} - 2 \int_0^{x_2} \frac{dt\; t}{y} + c\\
&=\, \frac{\omega_1}{c_4} \int_0^{\bar{x}_2} \Psi_0(0,t,\vec{b}) -2 \left[\int_0^{\bar{x}_2} \left(\Psi_{-1}(\infty,t,\vec{b})\right)+ (b_- + 2 c_4 G_\ast(\vec{b})) \frac{\omega_1}{c_4} \Psi_0(0,t,\vec{b})\right] + c\\
&=\, -2\, \cEfe{-1}{\infty}{\bar{x}_2}{\vec{b}}\ .
\end{split}
\end{align}
In the expression above, the terms proportional $\cEfe{0}{0}{\bar{x}_2}{\vec{b}}$ cancel out since for the elliptic curve under consideration $G_\ast(\vec{b})$ and $c_4$ are related and in particular $G_\ast(\vec{b})$ evaluates to a simple algebraic function, 
\begin{align}
\label{eq:Gstar_ttbar}
\begin{split}
c_4 \,&=\, \frac{1}{4}\left(1-\sqrt{1-16 a}\right)\,,\\
G_\ast(\vec{b}) \,&=\, -1+\frac{1}{1-\sqrt{1-16 a}}\, ,
\end{split}
\end{align}
see eq.~\eqref{eq:Gstarrat}.
Moreover, we fixed the boundary term $c=0$ by expanding $\Gm(\bar{x}_2) $ around $\bar{x}_2=0$.
Performing similar steps on all MPLs appearing in the integrand of eq.~\eqref{eq:last_int} we are able to express 
the integral as eMPLs, obtaining a very compact expression in terms of a weight-one prefactor times a pure combination of eMPLs of uniform weight 3,
\begin{align}
\begin{split}
\label{eq:M1}
M_1 \,=\, & \Omega_1^{(t\bar{t})}  \tilde{M_1}\ ,
\end{split}
\end{align}
with
\begin{align}
\label{eq:integral_final_v1}
\begin{split}
\Omega_1^{(t\bar{t})}\,=\,& -\frac{16\, a^2\, \omega _1}{m^4(1-\sqrt{1-16 a})}\, \quad \tilde{M}_1 \,=\, 5 \, T_{1+}(a) + 3\, T_{1-}(a) + \mathcal{O}(\epsilon)\ ,\\[5pt]
T_{1+}(a)\,=\,&\cEf{0&-1&1&1}{0&\infty&0&1-\rmp}{1}+\cEf{0&-1&1&1}{0&\infty&0&\rmp}{1}+\cEf{0&-1&1&1}{0&\infty&1&1-\rmp}{1}+\cEf{0&-1&1&1}{0&\infty&1&\rmp}{1}\ ,\\[5pt]
T_{1-}(a)\,=\,&-\cEf{0&-1&1&1}{0&\infty &\rmm&0}{1}-\cEf{0&-1&1&1}{0&\infty &\rmm&1}{1}-\cEf{0&-1&1&1}{0&\infty &1-\rmm&0}{1}-\cEf{0&-1&1&1}{0&\infty &1-\rmm &1}{1}\\[5pt]
&+\log (a) \left[\cEf{0&-1&1}{0&\infty&1-\rmm}{1}+\cEf{0&-1&1}{0&\infty&\rmm}{1}\right]\ ,
\end{split}
\end{align}
where we omitted the dependence of $\mathcal{E}_4$ on the branch points $\vec{b}$ for clarity and restored the factors of $m^2$ using dimensional analysis.
Note that, in order to obtain the expression above, we used the fact that for this region the expression 
for $G_*(\vec{b})$ reduces to the simple algebraic function in eq.~\eqref{eq:Gstar_ttbar}.
We stress once more that a Feynman integral which can be put in this form 
(i.e. one single prefactor which depends on the elliptic periods times a combination of elliptic polylogarithms
of uniform transcendental weight) 
appears to be the natural generalisation of a \textsl{pure integral} from the polylogarithmic to the elliptic case.

We remind the reader that the expression in eq.~\eqref{eq:integral_final_v1} is valid in the region 
where $a>1/16$ for which two of the roots of the elliptic curve are located outside of the real axis, 
and the branch points are ordered according to the conventions outlined in eq.~\eqref{eq:two_complex_roots}, as shown in eq.~\eqref{eq:rootsttbar}.
It is easy to see that in the Euclidean region $0 < a < 1/16$ the roots are real and can be ordered on the real axis between 0 and 1 (see fig.~\ref{fig:punctures} below). Moreover, the special points $r_{+/-}$ defined in eq.~\eqref{eq:rs} also become real and, in particular, one finds 
$0<r_-<1$.
Therefore, in this region the pole $r_-$ lies on the integration contour and eq.~\eqref{eq:integral_final_v1} is not a convenient representation.
As an example of the analytic continuation of an expression written in terms of eMPLs, in Appendix~\ref{sec:top-empl} we show how to analytically continue eq.~\eqref{eq:integral_final_v1} to the region $0 < a < 1/16$.
\begin{figure}[h]
\centering
\includegraphics[width=0.7\linewidth]{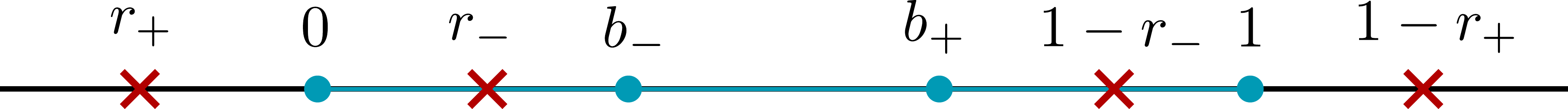}
\caption{Location of the ranch points of the final expression for the top-production triangle integral~\eqref{eq:integral_final_v1} in terms of eMPLs $\text{E}_4$ in the region $0<a<1/16$.}
\label{fig:punctures}
\end{figure}

\subsection{Second master integral}
\label{sec:ttbar-second-master}

Performing similar steps for the second master integral, $M_2$ in eq.~\eqref{eq:mastersttbar}, we can express it 
in terms of the same class of functions.
Contrary to the first master, we find that the second master integral 
cannot be written as one single (transcendental) prefactor
times a pure combination eMPLs of uniform weight.
Instead, by direct integration we find 
\begin{equation}
\label{eq:M2}
M_2 = \Omega_2^{(t\bar{t})} \, \tilde{M}_1 + H_2^{(t\bar{t})} \tilde{M}_2\,,
\end{equation}
where 
\beq
\bsp
\label{eq:Om2H2}
\Omega_2^{(t\bar{t})}\,&=\, \frac{\left(1+\sqrt{1-16 a}\right) (1-20 a)\, a}{12 m^4 (1-16 a)} \omega _1 + \frac{2\left(1-\sqrt{1-16 a}\right)\, a^2}{m^4(1-16 a)}  \eta _1 + \ord(\eps)\ ,\\[5pt]
H_2^{(t\bar{t})} \,&=\, -\frac{\left(1-\sqrt{1-16 a}\right) a^2}{m^4 (1-16 a)} \frac{1}{ \omega _1} + \ord(\eps)\,,
\esp
\eeq
and $\tilde{M}_1$ is the very same combination of eMPLs defined in eq.~\eqref{eq:integral_final_v1}. 
$\tilde{M}_2$ is instead another (independent) \textsl{pure combination} of eMPLs very similar to eq.~\eqref{eq:integral_final_v1} 
given by
\beq
\bsp
\label{eq:M2tilde}
\tilde{M_2} \,=\, 5 \, T_{2+}(a) + 3\, T_{2-}(a)+ \mathcal{O}(\epsilon)\ ,
\esp
\eeq
where $T_{2+}(a)$ and $T_{2-}(a)$ are pure functions of uniform weight four given by
\begin{align}
\begin{split}
T_{2+}(a)\,=\,&\cEf{-2&1&1}{\infty &0&\rmp}{1}+ \cEf{-2&1&1}{\infty &1&\rmp}{1}+ \cEf{-2&1&1}{\infty &0&\rpp}{1}+ \cEf{-2&1&1}{\infty &1&\rpp}{1}\ ,\\[5pt]
T_{2-}(a)\,=\,&-\cEf{-2&1&1}{\infty &\rmm&0}{1}-\cEf{-2&1&1}{\infty &\rmm&1}{1}-\cEf{-2&1&1}{\infty &1-\rmm&0}{1}-\cEf{-2&1&1}{\infty &1-\rmm&1}{1}\\
& + \log (a) \left[\cEf{-2&1}{\infty &\rmm}{1}+\cEf{-2&1}{\infty &\rpm}{1}\right]\ .
\end{split}
\end{align}
We note here that, according to the prescription provided in ref.~\cite{Broedel:2018qkq},
the prefactors $\Omega_2^{(t\bar{t})}$ and $H_2^{(t\bar{t})}$ in formula~\eqref{eq:M2}  have different
``weights'', namely $\Omega_2^{(t\bar{t})}$ has weight $w_{\Omega_2^{(t\bar{t})}}=1$, while $H_2^{(t\bar{t})}$ has weight $w_{H_2^{(t\bar{t})}}=-1$.
Similarly, while the combination $\tilde{M}_1$ has weight $w_{\tilde{M}_1} = 3$, $\tilde{M}_2$ has weight 
$w_{\tilde{M}_2}=4$. This shows that the second master integral $M_2$, as it stands, is not a function
of uniform transcendental weight.

Nevertheless, direct inspection of eq.~\eqref{eq:M2} 
suggests that we perform a change of basis from the master integrals $M_1$, $M_2$ 
to a new basis of pure master integrals
$\tilde{M}_1$, $\tilde{M}_2$
defined as
\beq
\label{eq:ttbar_triangle}
\begin{pmatrix}M_1 \\ M_2 \end{pmatrix} \,=\, \begin{pmatrix}\Omega_1^{(t\bar{t})} & 0  \\ \Omega_2^{(t\bar{t})} & H_2^{(t\bar{t})}\end{pmatrix} \begin{pmatrix}\tilde{M}_1 \\ \tilde{M}_2
\end{pmatrix} \,.
\eeq

It turns out that the prefactors $\Omega_1^{(t\bar{t})}$ and $\Omega_2^{(t\bar{t})}$ have a natural interpretation in terms of  
the maximal cuts of the integrals $M_1$ and $M_2$. 
In order to see why this is the case, let us recall that the two master integrals $M_1$ and $M_2$ fulfil a system of two coupled 
differential equations which, neglecting the subtopologies, read~\cite{vonManteuffel:2017hms} 
\begin{align}
\partial_a \begin{pmatrix}M_1 \\ M_2 \end{pmatrix} =  \begin{pmatrix}\frac{2}{a} & \frac{4}{a} \\ \frac{1}{1-16 a} & \frac{2(1-8 a)}{a (1-16 a)}\end{pmatrix} \begin{pmatrix}M_1 \\ M_2 \end{pmatrix} 
=\mathcal{H} \,
\begin{pmatrix}M_1 \\ M_2 \end{pmatrix}\,. \label{eq:deqsttb}
\end{align}
A complete solution of eq.~\eqref{eq:deqsttb} can be obtained by considering the matrix
of the maximal cuts of the two master integrals evaluated along two independent integration contours~\cite{Primo:2016ebd,Bosma:2017ens,Primo:2017ipr,Harley:2017qut}.
The number of independent contours is always equal to the number of master integrals. We denote a basis of independent contour by $\cC_j$, $j=1,2$.
By indicating the 
maximal cut of the integral $M_i$ along the contour $\mathcal{C}_j$ as ${\rm Cut}_j(M_i)$ we find
\begin{align}
G = \begin{pmatrix} {\rm Cut}_1(M_1) & {\rm Cut}_2(M_1) \\ {\rm Cut}_1(M_2) & {\rm Cut}_2(M_2)  \end{pmatrix}\, 
\quad \Rightarrow \quad
\partial_a G =  \mathcal{H} G\,.
\end{align}
This in turn implies that if we define a new basis of master integrals as
\begin{align}
 \begin{pmatrix}M_1 \\ M_2 \end{pmatrix} = G  \begin{pmatrix}F_1 \\ F_2 \end{pmatrix} \,,
\end{align}
then by construction the new basis fulfils ${\rm Cut}_j{(F_i)} = \delta_{ij}$\,.
Let us write ${\rm Cut}_1(M_j) = \overline{\Omega}_j$ and ${\rm Cut}_2(M_j) = \overline{H}_j$, and decompose 
the matrix $G$ as follows
\begin{align}
G =  \begin{pmatrix} \overline{\Omega}_1 & \overline{H}_1 \\  \overline{\Omega}_2 & \overline{H}_2 \end{pmatrix}
= \begin{pmatrix} \overline{\Omega}_1 & 0 \\  \overline{\Omega}_2 & X_2 \end{pmatrix}
\begin{pmatrix} 1 & \tau \\ 0 & 1 \end{pmatrix}\, \quad \mbox{with}\;\;\; \tau = \frac{\overline{\Omega}_1}{\overline{H}_1}\,, \;\;
X_2 = \frac{\overline{\Omega}_1\overline{H}_2 - \overline{\Omega}_2 \overline{H}_1}{\overline{\Omega}_1}\,. \label{eq:deccuts}
\end{align}
Comparing eq.~\eqref{eq:deccuts} with eq.~\eqref{eq:ttbar_triangle}, one can explicitly make the identifications
\begin{align}
\begin{pmatrix} 1 & \tau \\ 0 & 1 \end{pmatrix}  \begin{pmatrix}F_1 \\ F_2 \end{pmatrix} = \begin{pmatrix}\tilde{M}_1 \\ \tilde{M}_2 \end{pmatrix}\,,
\quad \begin{pmatrix} \overline{\Omega}_1 & 0 \\  \overline{\Omega}_2 & X_2 \end{pmatrix} = \begin{pmatrix}\Omega_1^{(t\bar{t})} & 0  \\ \Omega_2^{(t\bar{t})} & H_2^{(t\bar{t})} \end{pmatrix}\,, 
\end{align}
such that $\Omega_1^{(t\bar{t})}$ and $\Omega_2^{(t\bar{t})}$ correspond to the maximal cuts of the two master integrals along the first integration contour.
Clearly, the choice of which integration contour is considered to be the first one and which the second ones 
is arbitrary. This is
reflected in the ambiguity of the splitting in eq.~\eqref{eq:deccuts}.
Using this insight, we can verify that indeed $\Omega_1^{(t\bar{t})}$ and $\Omega_2^{(t\bar{t})}$ fulfil the differential equation in eq.~\eqref{eq:deqsttb}
\beq\bsp
\label{eq:de_maxcut_ttbar}
\Omega_2^{(t\bar{t})} \,=\,   \frac{1}{4}\left(a \partial_a - 2\right)\Omega_1^{(t\bar{t})}\ .
\esp\eeq

Before summarising the main points of these calculations, 
it is worth stressing once again that the result that we obtained, in particular in terms of two independent combinations of pure functions eq.~\eqref{eq:ttbar_triangle},
 is a-priori non trivial and it provides a strong hint towards the generalisation
of the idea of a pure basis of master integrals to the elliptic case.

\subsection{Summary}

With the explicit computation of the family of non-planar triangles considered in this section, we have discovered an elegant structure underlying these integrals. In particular, we highlight here the main features of the computation and result:
\begin{itemize}
\item eMPLs provide a natural language to express Feynman integrals which depend on an elliptic curve. Their recursive definition~\eqref{eq:cE4_def} provides an algorithmic method for computing each integration step. In the examples considered in this section, we applied the Cheng-Wu theorem in order to delay the appearance of the square root defining the 
elliptic curve to the last integration, but we stress that this is only for technical simplicity, 
the framework can accommodate a dependence on the elliptic curve at any step. The only caveat is that, as mentioned earlier, the framework is suited for cases where only a single elliptic curve is present, and thus in earlier integrations steps one cannot handle elliptic curves that still depend on other integration variables.
\item The results of the two master integrals shown in eqs.~\eqref{eq:M1} and \eqref{eq:M2} 
are naturally organised in terms of pure building blocks of uniform weight. 
These results can easily be rewritten in a new basis of pure master integrals. 
However, in contrast with the polylogarithmic case, the change of basis is \textsl{not} 
algebraic and depends on the periods and quasi-periods of the elliptic curve 
(see eqs.~\eqref{eq:integral_final_v1} and \eqref{eq:Om2H2}).
\item Since the Feynman-parameter integrals we started from are simply integrals over rational functions, 
the final result expressed in terms of eMPLs should reflect the property that the original integral 
has no intrinsic dependence on which sign we choose for the square root $y$. As such, 
the final integral should be invariant under the parity transformation $y \rightarrow -y $. Using the fact that $\omega_1,\,\eta_1$ and $Z_4(x,\vec{a})$ 
are parity odd, it is easy to see from the definitions of the eMPLs kernels in 
eqs.~\eqref{eq:pure_psi0},  \eqref{eq:pure_psi1},  \eqref{eq:pure_psi2} that under $y \rightarrow -y $ the kernels transform as
\beq\bsp
\Psi_0(x) \rightarrow \Psi_0(x)\,,\quad \Psi_{\pm n} (x) \rightarrow \pm \Psi_{\pm n} (x) \,  \quad n>0\ .
\esp\eeq
With this, one can see that every term in the final results for the two master integrals are manifestly parity-even as desired.
\end{itemize}
After learning the details of this computation and the properties of the result, we now turn our attention to a different integral which has a very similar structure: a particular contribution to the electroweak form factor.


\section{Electroweak form factor}
\label{sec:ewff}
\begin{figure}[h]
	\centering
	\includegraphics[width=0.3\linewidth]{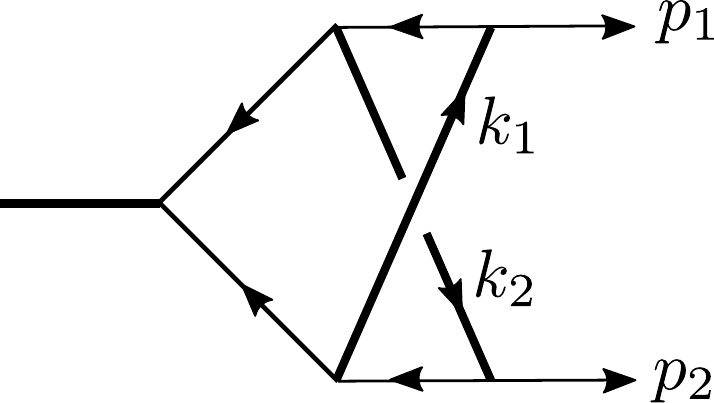}
	\caption{Triangle with massive loop.}
	\label{fig:trimassloop-ewff}
\end{figure}

In this section we consider another two-loop three-point function which is at first sight very similar to the one considered in the previous section. 
The topology is shown in fig.~\ref{fig:trimassloop-ewff} and differs from that of fig.~\ref{fig:trimassloop} in that 
two additional internal propagators are massless.

This family of integrals contributes to non-planar two-loop corrections to the electroweak form factor and was 
studied in ref.~\cite{Aglietti:2004tq}, where an approximate solution as series expansions around all singular points was obtained. 
We define the integral family as
\begin{align}
\label{eq:ewff-int}
J_{a_1,\dots,a_7}\,=\,\int\! \frac{d^D k_1 d^D k_2}{(i \pi)^{D}} \frac{D_7^{-a_7}}{\prod_{i=1}^6 D_i^{a_i}}\ ,
\end{align}
where as before $a_7 < 0$ and the remaining $a_i > 0$ are the powers of the propagators below,
\begin{align}
\begin{split}
D_1\,&=\, k_1^2-m^2, \quad D_3\,=\,(k_1-p_1)^2,\quad D_5\,=\,(k_1-k_2-p_1)^2 , \\
 D_2\,&=\,k_2^2-m^2, \quad  D_4\,=\, (k_2-p_2)^2,\quad D_6\,=\,(k_2-k_1-p_2)^2,\quad 
D_7 = k_1 \cdot p_2\,.
\end{split}
\end{align}
As for the previous family of integrals, we define $q^2 = (p_1+p_2)^2 = -m^2/a$, where $a$ is a dimensionless
ratio.
In contrast with the integrals discussed in Section~\ref{sec:ttbar}, 
in this case the top-sector is reduced to  three master integrals
which satisfy a coupled system of three differential equations. 
We choose the three master integrals as follows
\begin{align}
\label{eq:ewff-masters}
N_1\,=\,J_{1,1,1,1,1,1,0}\,,\quad N_2\,=\,J_{2,1,1,1,1,1,0} \,,\quad N_3\,=\,J_{1,1,1,1,1,1,-1}\ ,
\end{align}
where we notice that all three master integrals are finite in $D=4$ space-time dimensions.

We proceed similarly to the previous example and integrate all three master integrals explicitly starting from their 
Feynman parameter representation. 
As before, in order to be able to express all three master integrals in terms eMPLs, 
we need to find an ordering of Feynman parameters (or more precisely an application of the Cheng-Wu theorem)
which allows us to perform all integrations either in terms of standard MPLs or by introducing at most one square
root of a quartic polynomial in one of the integration variables. 
This square root will define the elliptic curve associated to the problem.
By direct inspection we find that the same application of the Cheng-Wu theorem as in the previous section does the trick. The polynomial equation defining the elliptic curve resulting from it, though, appears to be substantially
more complicated, at least once its roots are seen as functions of the kinematic invariant $a$ defined in eq.~\eqref{eq:adef} above.
The elliptic curve is defined by
\begin{align}
\label{eq:ewffellcurve}
y^2 = (x-d_-)(x-d_+)(x-1+d_-)(x-1+d_+) \ ,
\end{align}
where the roots are defined as
\beq
\label{eq:dpm}
d_\pm \,=\,\frac{1}{2} \left(1-\sqrt{1-4 a (1+2 a)\pm8 \sqrt{a^3 (a+1)}}\right)\, .
\eeq
Moreover, in the intermediate integration steps we find standard MPLs which have branch cuts in the special points
$x \in \left\{ a,-a,1+a,1-a,r_-,1-r_- \right\}$ with
\beq
r_- = \frac{1}{2}(1-\sqrt{1-4a})\,.
\eeq 

Similarly to the integrals considered in the previous section, since the integration over the last Feynman parameter varies 
in the interval $(0,1)$,
it is convenient to work in a kinematical region where there are no 
explicit poles on their integration contour. 
Since, in order to implement Feynman's prescription, the dimensionless ratio $a$ becomes
a complex number with a small negative imaginary part,
we choose to work with $\text{Re}(a)>1$, such that the all poles lie outside of the 
integration contour and, in particular,
$r_{+/-}$ are complex conjugate to each other. 
In this situation, two of the branch points of the elliptic curve in eq.~\eqref{eq:ewffellcurve} are real and located between 0 and 1, 
whereas the two remaining branch points are complex conjugate to each other and have real part equal to $1/2$. 
We order the branch points of the elliptic curve according to the conventions of Section~\ref{sec:eMPLs} such that 
eqs.~\eqref{eq:two_complex_roots} and \eqref{eq:relationstwocomplex} are satisfied, namely
\begin{align}
\begin{split}
\vec{d}\,=\, (d_-, 1-d_+, d_+, 1-d_-)\,,
\end{split}
\end{align}
with $d_{\pm}$ given in eq.~\eqref{eq:dpm}.
In order to arrive at compact expressions for the integrals in terms of pure eMPLs, we use the fact that the functions $c_4(\vec{d}),\, Z_4(x, \vec{d})$ and $G_*(\vec d)$ admit particularly simple representations in this case, namely
\beq\bsp
c_4(\vec{d})\,=\, & -\frac{1}{2}\,d_{+-} \ , \\
Z_4(0,\vec{d})\,=\, & -\frac{1+4a}{3 \,d_{+-}} + \frac{8 \pi i}{3 \omega_1}\ ,\\
Z_4(1,\vec{d})\,=\, & \frac{1+4a}{3 \,d_{+-}} + \frac{4 \pi i}{3 \omega_1}\ ,\\
G_*(\vec{d})\,=\, & \frac{2 d_- -1}{2\,d_{+-}}\, ,
\esp\eeq
where we defined $d_{+-} = d_+ - d_-$.
While the evaluation of the three master integrals proceeds at least conceptually along 
the same lines described in detail in Section~\ref{sec:ttbar},
the individual manipulations and the final results are more cumbersome, mainly due to the 
explicit form of the branch points of the elliptic curve.

The result for the first master integral in terms of pure eMPLs is given by 
\beq\bsp
N_1\,=\, \Omega_1^{\rm (ew)} \tilde{N}_1\ , 
\esp\eeq
where
\begin{align}
 \Omega_1^{\rm (ew)} \,=\,&- \frac{2 a^2}{m^4 \,d_{+-}}\,  \omega _1\,,\quad \tilde{N}_1\,=\,2 Q_{1-}(a) + Q_1(a) + \mathcal{O}(\epsilon)\ ,
 \nonumber \\[5pt]
Q_{1-}(a)\,&=\,\cEf{0&-1&1&1}{0& 0 & \rmm & 0}{1}+\cEf{0&-1&1&1}{0& 0 & \rmm & 1}{1}+\cEf{0&-1&1&1}{0&1&\rmm&0}{1}+\cEf{0&-1&1&1}{0& 1 & \rmm & 1}{1}  \nonumber \\
& + 2\Big[\cEf{0&-1&1&1}{0& \infty & \rmm & 0}{1}+\cEf{0&-1&1&1}{0& \infty & \rmm & 1}{1}\Big] -4 i \pi \Big[\cEf{0&0&1&1}{0& 0 & \rmm & 0}{1}+ \cEf{0&0&1&1}{0& 0 & \rmm & 1}{1}\Big]   \nonumber\\
&-\log (a)\Big[2 \cEf{0&-1&1}{0&\infty &\rmm}{1}+\cEf{0&-1&1}{0&0&\rmm}{1}+\cEf{0&-1&1}{0&1&\rmm}{1}-4 i \pi  \cEf{0&0&1}{0&0&\rmm}{1}\Big]  \nonumber\\
& + (\rmm \rightarrow 1-\rmm)\ ,  \nonumber \\[5pt]
Q_1(a)\,=\,& -4 \Big[\cEf{0&-1&1&1}{0&\infty & 0& -a}{1}+\cEf{0&-1&1&1}{0&\infty & 1 & 1+a}{1}
+\cEf{0&-1&1&1}{0&\infty & 0 & 1}{1}+\cEf{0&-1&1&1}{0 &\infty & 1 & 0}{1}\Big]  \nonumber\\
   &-3 \Big[ \cEf{0&-1&1&1}{0 &\infty & 1-a & 1}{1}+\cEf{0&-1&1&1}{0 &\infty & 1-a & -a}{1}+\cEf{0&-1&1&1}{0 &\infty & a & 0}{1}+\cEf{0&-1&1&1}{0 &\infty & a & 1+a}{1}  \nonumber\\
   &+\cEf{0&-1&1&1}{0 & 0 & a & 0}{1}+\cEf{0&-1&1&1}{0 & 0 & a & 1+a}{1}+\cEf{0&-1&1&1}{0 & 1 & 1-a & 1}{1}+\cEf{0&-1&1&1}{0 & 1 & 1-a & -a}{1}\Big]  \nonumber \\
   &+\cEf{0&-1&1&1}{0 & 0 & 1+a & 1}{1}+\cEf{0&-1&1&1}{0 & 1 & -a & 0}{1}-2
   \Big[ \cEf{0&-1&1&1}{0 & 0 & 0 & -a}{1}+\cEf{0&-1&1&1}{0 & 0 & 1 & 1+a}{1}  \nonumber \\
   &+\cEf{0&-1&1&1}{0 & 0 & -a & 0}{1}+\cEf{0&-1&1&1}{0 & 1 & 0 & -a}{1}+\cEf{0&-1&1&1}{0 & 1& 1 & a+1}{1}+\cEf{0&-1&1&1}{0 & 1 & 1+a & 1}{1}\Big]  \nonumber \\
   &+3
   (\cEf{0&-1&1&1}{0 & \infty & -a & 0}{1}+\cEf{0&-1&1&1}{0 & \infty & 1+a & 1}{1})+4 i \pi  \Big[ \cEf{0&0&1&1}{0 & 0 & 1-a & 1}{1}+\cEf{0&0&1&1}{0 & 0 & 1-a & -a}{1}  \nonumber\\
    &+\cEf{0&0&1&1}{0 & 0 & -a & 0}{1}+2 
   \cEf{0&0&1&1}{0 & 0 & 0 & -a}{1}+2\cEf{0&0&1&1}{0 & 0 & 1 & 1+a}{1}+2\cEf{0&0&1&1}{0 & 0 & a & 0}{1}  \nonumber \\
   &+2\cEf{0&0&1&1}{0 & 0 & a & 1+a}{1}\Big]+ \log \left(\frac{a+1}{a}\right) \Big[-2
   (\cEf{0&-1&1}{0&0&1}{1}+\cEf{0&-1&1}{0 & 1 & 1}{1})  \nonumber \\
   &+8 i \pi  (\cEf{0&0&1}{0 & 0 & a}{1}+\cEf{0&0&1}{0&0&1}{1}) 
     -4 \cEf{0&-1&1}{0 & \infty & 1}{1}-3 (\cEf{0&-1&1}{0 & \infty & a}{1}+\cEf{0&-1&1}{0 & 0 & a}{1})\Big]  \nonumber \\
   & + \frac{1}{6} \cEf{0}{0}{1} \left(-6 G(-1,0,-1,a)+6 G(0,0,-1,a)-\pi ^2 \log \left(\frac{a+1}{a}\right)-12 \zeta_3 \right)\ .
  \end{align}

The three master integrals in eq.~\eqref{eq:ewff-masters} corresponding to the top-dimensional topologies of this family of integrals follow a structure similar to that observed in the previous section for the triangle with a massive loop. Indeed, after computing the second and third master integrals also through direct integration of the Feynman-parametric integral, it is possible to transform 
them into a basis of functions which are pure combinations of eMPLs of uniform weight,
\beq
\label{eq:ewff_triangle}
\begin{pmatrix}N_1 \\ N_2 \\ N_3\end{pmatrix} \,=\, \begin{pmatrix}\Omega_1^{\rm (ew)}  & 0 & 0 \\ \Omega_2^{\rm (ew)}  & H_2^{\rm (ew)}  & 0\\ \Omega_3^{\rm (ew)}  & 0 & X_3^{\rm (ew)}  \end{pmatrix} \begin{pmatrix}\tilde{N}_1 \\ \tilde{N}_2 \\ \tilde{N}_3 \end{pmatrix} \ .
\eeq
The entries of the matrix in eq.~\eqref{eq:ewff_triangle} are given by
\beq\bsp
\Omega_2^{\rm (ew)} \,&=\, \frac{a^2 (4 a (3 a+2)-1)}{2 m^4 (8 a-1) (a+1) \, d_{+-}}  \, \omega_1\  + 3 \frac{a^2 \, d_{+-} }{m^4 (1+a)(8a - 1)} \, \eta_1 \\
H_2^{\rm (ew)} \,&=\, \frac{a^2 d_{+-}}{ 6 m^4 (1+a)(1-8 a)} \frac{1}{\omega_1}\ ,\quad
\Omega_3^{\rm (ew)} \,=\,  \frac{(2-a)a}{3 m^2 d_{+-}} \omega_1 \,,\quad X_3^{\rm (ew)}  \,=\, \frac{a}{72\, m^2}\,.
\esp\eeq

Similarly to the triangle considered in the previous section, the $\Omega_{i}^{\rm (ew)} $ above satisfy the differential equations for the maximal cuts of the three master integrals (see eq.~\eqref{eq:de_maxcut_ttbar}), namely:
\beq\bsp
\Omega_2^{\rm (ew)}  \,=&\, \frac{1}{2}(a\partial_a -2)\Omega_1^{\rm (ew)} \ , \\
(1-a \partial_a)\Omega_3^{\rm (ew)}  \,=&\, \frac{1}{6a}\left[(2-a)a \partial_a - (4-a)\right] \Omega_1^{\rm (ew)}  \ .
\esp\eeq

The pure function $\tilde{N}_2$ is a weight-four function which we can decompose into a part that depends on the variable $\rmm$ as well as a part with dependence only on the variable $a$ and a piece which is purely polylogarithmic. It is given by
\begin{align}
\tilde{N}_2 \,=\, &18 Q_{2-}(a) + 9 Q_2(a) + Q_{2,\text{MPL}}(a) + \mathcal{O}(\epsilon) \,,  \nonumber\\[5pt]
Q_{2-}(a)\,&=\,-2 \cEf{-2&1&1}{\infty &\rmm&0}{1}-2 \cEf{-2&1&1}{\infty &\rmm&1}{1}
-\cEf{-2&1&1}{0&\rmm&0}{1}-\cEf{-2&1&1}{0&\rmm&1}{1}  \nonumber\\
&+\cEf{1&-2&1}{\rmm&1&0}{1}+\cEf{1&-2&1}{\rmm&1&1}{1}+\cEf{1&1&-2}{\rmm&0&1}{1}+\cEf{1&1&-2}{\rmm&1&1}{1} \nonumber \\ 
&2 i \pi  \left[-\log (a) \cEf{1&1}{\infty &\rmm}{1}+\cEf{1&1&1}{\infty &\rmm&0}{1}+\cEf{1&1&1}{\infty &\rmm&1}{1}\right] \nonumber\\
&+\log (a) \left[ 2\cEf{-2&1}{\infty & \rmm}{1} + \cEf{-2&1}{0&\rmm}{1}-\cEf{1&-2}{\rmm&1}{1}\right] + (\rmm \rightarrow 1-\rmm)\,,
 \nonumber\\[5pt]
Q_2(a)\,&=\, 4 \cEf{-2&1&1}{\infty &0&-a}{1}+4 \cEf{-2&1&1}{\infty &1&a+1}{1}
+3 \cEf{-2&1&1}{\infty &1-a&1}{1}+3 \cEf{-2&1&1}{\infty &1-a&-a}{1}  \nonumber \\
&-3 \cEf{-2&1&1}{\infty &-a&0}{1}+3 \cEf{-2&1&1}{\infty &a&0}{1}+3 \cEf{-2&1&1}{\infty &a&a+1}{1}
-3 \cEf{-2&1&1}{\infty &a+1&1}{1} \nonumber \\
&+\log \left(\frac{1}{a}+1\right) (3 \cEf{-2&1}{\infty &a}{1}+3 \cEf{-2&1}{0&a}{1}+4 \cEf{-2&1}{\infty &1}{1}
+2 \cEf{-2&1}{0&1}{1})  \nonumber \\
&+2 \cEf{-2&1&1}{0&0&-a}{1}+2 \cEf{-2&1&1}{0&1&a+1}{1}+2 \cEf{-2&1&1}{0&-a&0}{1}
+3 \cEf{-2&1&1}{0&a&0}{1}  \nonumber \\
&+3 \cEf{-2&1&1}{0&a&a+1}{1}-\cEf{-2&1&1}{0&a+1&1}{1}-2 \cEf{1&-2&1}{0&1&-a}{1}-3 \cEf{1&-2&1}{1-a&1&1}{1} \nonumber \\
&-3 \cEf{1&-2&1}{1-a&1&-a}{1}+\cEf{1&-2&1}{-a&1&0}{1}-2 \cEf{1&-2&1}{a+1&1&1}{1}-2 \cEf{1&1&-2}{0&-a&1}{1} \nonumber \\
&-3 \cEf{1&1&-2}{1-a&1&1}{1}-3 \cEf{1&1&-2}{1-a&-a&1}{1}+\cEf{1&1&-2}{-a&0&1}{1}+4 \cEf{-2&1&1}{\infty &0&1}{1} \nonumber\\
&+4 \cEf{-2&1&1}{\infty &1&0}{1}+4 \text {Li} _ 2 (-a) \cEf{-2}{\infty }{1}-4 \zeta_2\cEf{-2}{\infty }{1}-2 i \pi  \Big[ 2 \cEf{1&1&1}{\infty &0&-a}{1} \nonumber\\
&+2 \cEf{1&1&1}{\infty &1&a+1}{1}+\cEf{1&1&1}{\infty &1-a&1}{1}
+\cEf{1&1&1}{\infty &1-a&-a}{1}+\cEf{1&1&1}{\infty &-a&0}{1} \nonumber \\
&+2 \cEf{1&1&1}{\infty &a&0}{1}+2 \cEf{1&1&1}{\infty &a&a+1}{1}
+2 \log \left(\frac{1}{a}+1\right) \big(\cEf{1&1}{\infty &1}{1}+ \cEf{1&1}{\infty &a}{1}\big)\Big]\ , \nonumber\\
Q_{2,\text{MPL}}(a)\,&=\, 
i \pi ^3 \Big[15 \log (a)-3 \log (a+1)-4 \log \left((1-\rmm) \rmm^2\right)\Big]  \nonumber \\ 
&+12 \pi^2 \Big[\log (\rmm (1-\rmm)) \log (\rmm (1+\rmm))
-\log (a) \log \left((\rmm-1)(1+\rmm)\right)\Big]  \nonumber \\
&-i \pi  \Big[ 6 G_{-1,0,0}(a)+24 G_{0,0,1}(\rmm)+24 G_{1,1,0}(\rmm)
+6 \log (a) \log ^2\left(\frac{\rmm-1}{\rmm}\right) \nonumber \\
&-\log ^3(a)-4 \log ^3(1-\rmm)-4 \log ^3(\rmm)  -24 \zeta_3 \Big] \ .
\end{align}
Finally, the function $\tilde{N}_3$ is more complicated than the previous two and we prefer not to write it here explicitly.
Its expression is attached to the ancillary files of the arXiv submission of this paper.

As for the previous case of the $t\bar{t}$ triangle, we see that following eq.~\eqref{eq:ewff_triangle} the three elliptic 
Feynman master integrals
can be re-expressed in a new basis of three  pure master integrals, $\tilde{N}_1$, $\tilde{N}_2$ and $\tilde{N}_3$. 
This provides further evidence that it is possible to find a pure basis to represent master integrals which do not evaluate to MPLs.


\section{Kite with three distinct masses}
\label{sec:kite}
\begin{figure}[h]
\centering
\includegraphics[width=0.3\linewidth]{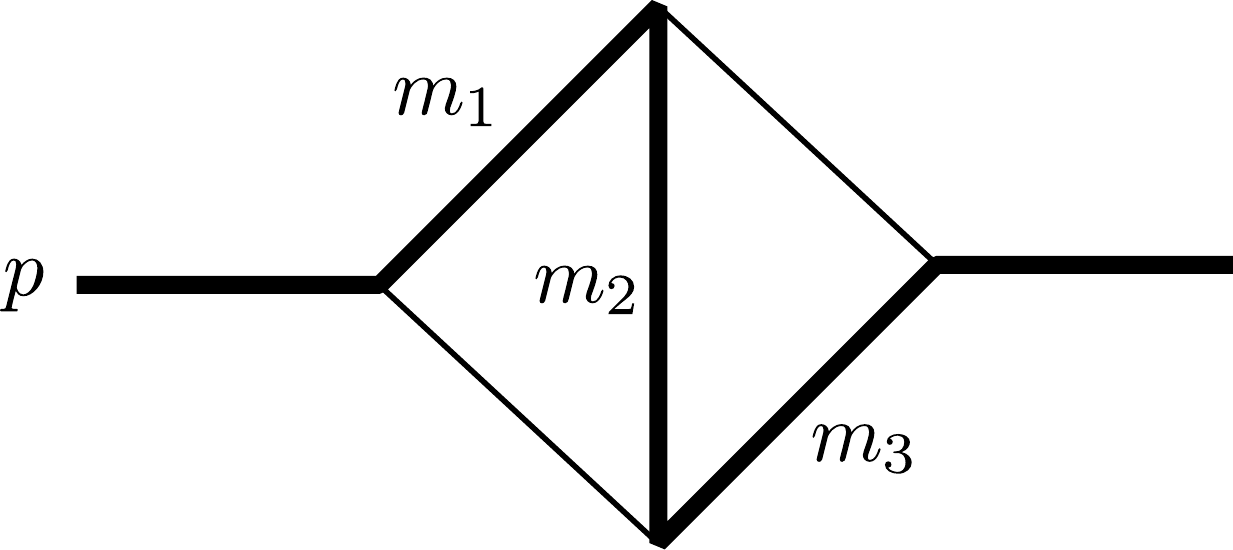}
\caption{Kite integral with three internal massive propagators with masses $m_1$, $m_2$ and $m_3$.}
\label{fig:kite3M}
\end{figure}

Having computed several three-point functions by performing the integrations over Feynman parameters in terms of pure eMPLs, we now consider an example of a two-point function with more scales, namely the kite integral with three distinct masses shown in fig.~\ref{fig:kite3M}.

\noindent The integral we consider is given by
\beq\bsp
K&(p^2,m_1^2,m_2^2,m_3^2)\\
 &= -\frac{e^{2\gamma_E\eps}}{\pi^D}\, \int \frac{d^Dk\, d^D l}{l^2\, (k-p)^2\, (k^2-m_1^2) ((k-l)^2-m_2^2) ((l-p)^2-m_3^2)}\,,
\label{eq:kite}
\esp\eeq
in $D=2-2\epsilon$. 
A simpler version of this integral, when all three internal masses have the same value $m_1=m_2=m_3=m$,
has been computed in the literature 
in terms of iterated integrals over products of elliptic integrals and polylogarithms~\cite{Remiddi:2016gno} or modular forms~\cite{Adams:2016xah}, in terms of elliptic generalisations of polylogarithms~\cite{Adams:2016xah} and finally, more recently, in terms of the
eMPLs considered here~\cite{Broedel:2018qkq}. We consider here the more general case with three different internal masses.
We encode the kinematic dependence is the three dimensionless ratios
\beq
a_i \,=\, -\frac{m_i^2}{p^2}\,,\quad i=1,2,3\ .
\eeq
We compute the kite integral in the region $0 < p^2 < \text{min}{(m_1^2,m_2^2,m_3^2)}$. The branch points are complex and given by
\begin{equation}
\vec{a} = \left\{ a_-,a_-^*, a_+, a_+^*\right\}\,,
\end{equation}
where
\begin{align*}\bsp
a_- \,&=\, \frac{\sigma_- +  \sqrt{\phi_-(\alpha_- + \beta_-)}}{2(1+a_3)^3}\,, \quad a_+ \,=\, \frac{\sigma_+ -i \sqrt{\phi_+(\alpha_+ + \beta_+)}}{2(1+a_3)^3}\, ,\\[5pt]
\sigma_+\,&=\, \left(\left(\sqrt{a_3}-i\right)^2-a_1+a_2\right) \left(\sqrt{a_3}-i\right) \left(\sqrt{a_3}+i\right)^3 \,,\\[15pt]
\sigma_-\,&=\, \left(\left(\sqrt{a_3}+i\right)^2-a_1+a_2\right) \left(\sqrt{a_3}-i\right)^3 \left(\sqrt{a_3}+i\right)\,,
\esp\end{align*}
\vspace{-13pt}
\begin{align}
\alpha_+\,&=\,-2 a_1 \left(a_2+\left(\sqrt{a_3}-i\right)^2\right)\,, &\alpha_-\,&=\, -2 a_1 \left(a_2+\left(\sqrt{a_3}+i\right)^2\right) \, ,\\[5pt]
\beta_+\,&=\, a_1^2+\left(a_2-\left(\sqrt{a_3}-i\right)^2\right)^2  \,,  &\beta_-\,&=\, a_1^2+\left(a_2-\left(\sqrt{a_3}+i\right)^2\right)^2\, ,\nonumber \\[5pt]
\phi_+\,&=\, -\left(\sqrt{a_3}-i\right)^2 \left(\sqrt{a_3}+i\right)^6\,, &\phi_-\,&=\, \left(\sqrt{a_3}-i\right)^6 \left(\sqrt{a_3}+i\right)^2 \nonumber \ .
\end{align}
As in the previous applications for three-point functions, the kite integral can be computed in terms of a pure combination of eMPLs of uniform weight three. In order to arrive at the final expressions, we make use of the following relations valid for the kinematic region we consider,
\beq\bsp
\label{eq:relations-kite}
G_\ast(\vec{a}) \,&=\, \frac{1}{2} \left(-Z_4\left(-\frac{a_2}{a_1-a_2},\vec{a}\right)+\frac{a_- \left(a_1-a_2\right)+a_2}{\left(a_2-a_1\right) c_4}+\frac{2 i \pi }{\omega _1}\right)\ ,\\[5pt]
Z_4\left(\frac{a_3-a_2}{a_1-a_2+a_3+1},\vec{a}\right)\,&=\, \frac{1}{\left(a_1-a_2\right) \left(a_3+1\right) \left(a_1-a_2+a_3+1\right) c_4}\Big[ c_4 Z_4(0,\vec{a}) \\
& +\left(a_1-a_2\right) \left(a_3+1\right) \left(a_1-a_2+a_3+1\right) c_4 Z_4\left(-\frac{a_2}{a_1-a_2},\vec{a}\right)\\
&-2 \left(a_1-a_2\right) \left(a_3+1\right) \left(a_1-a_2+a_3+1\right) +a_1^3-3 a_2 a_1^2+2 a_1^2\\
&+3 a_2^2 a_1-4 a_2 a_1+a_3 a_1+a_1-a_2^3+2 a_2^2+a_2 a_3^2+a_2 a_3\Big]+\frac{2 i \pi }{\omega _1}\ , \\[5pt]
Z_4(1,\vec{a}) \,&=\, \frac{-\left(a_3+1\right) c_4 Z_4(0,\vec{a})+a_1-a_2}{\left(a_3+1\right) c_4}+\frac{2 i \pi }{\omega _1}\ .
\esp\eeq
The result for the kite integral \eqref{eq:kite} with three distinct masses in terms of eMPLs is given by
\beq
K(p^2,a_i) =  \frac{1}{p^6}\Big[\tilde{K}(a_i)  +  \tilde{K}_{\rm MPL}(a_i)\Big] + \mathcal{O}(\epsilon)\ ,
\eeq
where
\begin{align}
\label{eq:kitecE4} 
\tilde{K}(a_i) \,=\, & \cEf{-1&-1&1}{0&0&1}{1}+\cEf{-1&-1&1}{0&1&1}{1}+\cEf{-1&-1&1}{0&\infty &1}{1}
-\cEf{-1&-1&1}{0&\lambda&1}{1}\nonumber \\
&+\cEf{-1&1&-1}{0&1&0}{1}+\cEf{-1&1&-1}{0&1&1}{1}+\cEf{-1&1&-1}{0&1&\infty }{1}
-\cEf{-1&1&-1}{0&1&\lambda}{1}\nonumber \\
&+\cEf{1&-1&-1}{0&1&0}{1}+\cEf{1&-1&-1}{0&1&1}{1}+\cEf{1&-1&-1}{0&1&\infty }{1}
-\cEf{1&-1&-1}{0&1&\lambda}{1} \nonumber \\
&-\cEf{1&-1&-1}{\rho&\infty &0}{1}-\cEf{1&-1&-1}{\rho&\infty &1}{1}-\cEf{1&-1&-1}{\rho&\infty &\infty }{1}
+\cEf{1&-1&-1}{\rho&\infty &\lambda}{1}\nonumber \\
&+\frac{1}{2}\Big[- \cEf{-1&-1&1}{\infty &0&1}{1}-\cEf{-1&-1&1}{\infty &1&1}{1}-\cEf{-1&-1&1}{\infty &\infty &1}{1}
+\cEf{-1&-1&1}{\infty &\lambda&1}{1}\nonumber \\
&+\cEf{-1&-1&1}{\xi&0&1}{1}+\cEf{-1&-1&1}{\xi&1&1}{1}+\cEf{-1&-1&1}{\xi&\infty &1}{1}
-\cEf{-1&-1&1}{\xi&\lambda&1}{1}\nonumber \\
&-\cEf{-1&1&-1}{\infty &1&0}{1}-\cEf{-1&1&-1}{\infty &1&1}{1}-\cEf{-1&1&-1}{\infty &1&\infty }{1}
+\cEf{-1&1&-1}{\infty &1&\lambda}{1}\nonumber \\
&+\cEf{-1&1&-1}{\xi&1&0}{1}+\cEf{-1&1&-1}{\xi&1&1}{1}+\cEf{-1&1&-1}{\xi&1&\infty }{1}
-\cEf{-1&1&-1}{\xi&1&\lambda}{1} \nonumber\\
&+\cEf{1&-1&-1}{0&\infty &0}{1}+\cEf{1&-1&-1}{0&\infty &1}{1}+\cEf{1&-1&-1}{0&\infty &\infty }{1}
-\cEf{1&-1&-1}{0&\infty &\lambda}{1}\nonumber \\
&-\cEf{1&-1&-1}{0&\xi&0}{1}-\cEf{1&-1&-1}{0&\xi&1}{1}-\cEf{1&-1&-1}{0&\xi&\infty }{1}+\cEf{1&-1&-1}{0&\xi&\lambda}{1}\nonumber \\
&+\cEf{1&-1&-1}{\rho&0&0}{1}+\cEf{1&-1&-1}{\rho&0&1}{1}+\cEf{1&-1&-1}{\rho&0&\infty }{1}-\cEf{1&-1&-1}{\rho&0&\lambda}{1}\nonumber \\
&-\cEf{1&-1&-1}{\rho&1&0}{1}-\cEf{1&-1&-1}{\rho&1&1}{1}-\cEf{1&-1&-1}{\rho&1&\infty }{1}+\cEf{1&-1&-1}{\rho&1&\lambda}{1}\nonumber \\
&+\cEf{1&-1&-1}{\rho&\lambda&0}{1}+\cEf{1&-1&-1}{\rho&\lambda&1}{1}+\cEf{1&-1&-1}{\rho&\lambda&\infty }{1}-\cEf{1&-1&-1}{\rho&\lambda&\lambda}{1}\nonumber \\
&+\cEf{1&-1&-1}{\rho&\xi&0}{1}+\cEf{1&-1&-1}{\rho&\xi&1}{1}+\cEf{1&-1&-1}{\rho&\xi&\infty }{1}-\cEf{1&-1&-1}{\rho&\xi&\lambda}{1}\Big]\nonumber \\
& -2 i \pi  \Big[ \cEf{0&-1&1}{0&0&1}{1}+  \cEf{0&-1&1}{0&1&1}{1}+  \cEf{0&-1&1}{0&\infty &1}{1}-  \cEf{0&-1&1}{0&\lambda&1}{1}\nonumber \\
&+  \cEf{0&1&-1}{0&1&0}{1}+  \cEf{0&1&-1}{0&1&1}{1}+  \cEf{0&1&-1}{0&1&\infty }{1}-  \cEf{0&1&-1}{0&1&\lambda}{1}\nonumber \\
&+  \cEf{1&0&-1}{\rho&0&0}{1}+  \cEf{1&0&-1}{\rho&0&1}{1}+  \cEf{1&0&-1}{\rho&0&\infty }{1}-  \cEf{1&0&-1}{\rho&0&\lambda}{1}\Big]\nonumber \\
&+ \log \left(\frac{a_3}{a_2}\right)\Big[ -\cEf{-1&1}{0&1}{1} +\frac{1}{2} \cEf{-1&1}{\infty &1}{1} -\frac{1}{2} \cEf{-1&1}{\xi&1}{1} -\cEf{1&-1}{0&1}{1}  \nonumber \\
&-\frac{1}{2} \cEf{1&-1}{0&\infty }{1} +2 i \pi  \cEf{0&1}{0&1}{1}+\frac{1}{2} \cEf{1&-1}{0&\xi}{1} -\frac{1}{2} \cEf{1&-1}{\rho&0}{1} +\frac{1}{2} \cEf{1&-1}{\rho&1}{1} \nonumber \\
& +\cEf{1&-1}{\rho&\infty }{1} -\frac{1}{2} \cEf{1&-1}{\rho&\lambda}{1} -\frac{1}{2} \cEf{1&-1}{\rho&\xi}{1} +2 i \pi  \cEf{1&0}{\rho&0}{1} \Big]\ ,
\end{align}
where we define the shorthand notation
\beq\bsp
&\xi \,=\, \frac{a_1 \left(a_3-a_2\right)}{a_1 \left(a_2 \left(a_3-1\right)+a_3\right)-a_2 a_3}\,,\quad \lambda \,=\, \frac{a_1}{a_1-a_2}\,,\quad \rho\,=\,\frac{1}{1-a_3}\,,\quad \eta \,=\, \frac{a_1 a_3}{a_1 + a_3 -a_1 a_3}\ .
\esp\eeq
The part of the result which depends only on ordinary MPLs, $ \tilde{K}_{\rm MPL}(a_i) $, in turn is given by

\begin{align*}
 &\tilde{K}_{\rm MPL}(a_i) \,=\, \GG\left(1,a_3\right) \Big[\GG\left(\eta ,0,a_2\right)-\GG\left(\eta ,a_3,a_2\right)-\GG\left(a_1,0,a_2\right)+\GG\left(a_1 a_3,a_3,a_2\right)\Big]\\
 &+\frac{3}{2} \GG\left(\eta ,0,0,a_2\right)-2 \GG\left(\eta ,0,1,a_2\right)-\GG\left(\eta ,\eta ,0,a_2\right)+\GG\left(\eta ,\eta ,a_3,a_2\right)\\
 &-\frac{1}{2} \GG\left(\eta ,a_1,0,a_2\right)+\GG\left(\eta ,a_1,1,a_2\right)+\frac{1}{2} \GG\left(\eta ,a_3,0,a_2\right)+\GG\left(\eta ,a_3,1,a_2\right)\\
 &-\GG\left(\eta ,a_3,a_3,a_2\right)-\frac{1}{2} \GG\left(\eta ,a_1 a_3,0,a_2\right)+\GG\left(a_3,\eta ,0,a_2\right)-\GG\left(a_3,\eta ,a_3,a_2\right)\\
 &+\frac{1}{4} \GG\left(0,a_2\right)^3-\frac{1}{4} \GG\left(0,a_3\right)^3+\frac{1}{6} \GG\left(1,a_1\right)^3+\frac{1}{6} \GG\left(a_3,a_2\right)^3\\
 &+\frac{1}{2} \GG\left(a_1 a_3,a_2\right) \Big[-\GG\left(1,a_1\right)^2+2 \GG\left(0,1,a_1\right)+\GG\left(1,0,a_1\right)\Big]\\
 &+\GG\left(0,0,1,a_1\right)+\GG\left(0,0,1,a_2\right)-\GG\left(0,0,a_3,a_2\right)+\frac{1}{2} \GG\left(0,1,0,a_1\right)-\frac{1}{2} \GG\left(0,1,0,a_3\right)\\
 &-\GG\left(0,1,1,a_1\right)-\GG\left(0,a_1,0,a_2\right)-\frac{1}{2} \GG\left(0,a_3,0,a_2\right)-\frac{1}{2} \GG\left(0,a_1 a_3,0,a_2\right)\\
 &+\GG\left(0,a_1 a_3,a_3,a_2\right)-\GG\left(1,0,1,a_1\right)-\frac{1}{2} \GG\left(1,1,0,a_1\right)-\frac{3}{2} \GG\left(a_1,0,0,a_2\right)\\
 &+\frac{3}{2} \GG\left(a_1,a_1,0,a_2\right)+\frac{1}{2} \GG\left(a_1,a_1 a_3,0,a_2\right)-\frac{3}{2} \GG\left(a_3,0,0,a_2\right)-\frac{1}{2} \GG\left(a_3,a_3,0,a_2\right)\\
 &+\frac{1}{2} \GG\left(a_3,a_1 a_3,0,a_2\right)+\GG\left(a_1 a_3,0,1,a_2\right)+\GG\left(a_1 a_3,0,a_3,a_2\right)-\GG\left(a_1 a_3,a_1,1,a_2\right)\\
 &+\frac{1}{2} \GG\left(a_1 a_3,a_3,0,a_2\right)-\GG\left(a_1 a_3,a_3,1,a_2\right)-\GG\left(a_1 a_3,a_1 a_3,a_3,a_2\right)+\frac{1}{4} \Big[\Big((\GG\left(0,a_2\right)\\
 &-\GG\left(0,a_3\right)+\GG\left(\eta ,a_2\right)-\GG\left(a_1,a_2\right)-\GG\left(a_3,a_2\right)\Big) \GG\left(0,a_1\right)^2+\Big(-2 \GG\left(0,a_2\right)^2\\
 &+2 \GG\left(0,a_3\right) \GG\left(0,a_2\right)+2 \GG\left(\eta ,a_2\right)^2-3 \GG\left(a_1,a_2\right)^2+2 \GG\left(a_3,a_2\right)^2\\
 &-4 \GG\left(1,a_3\right) \GG\left(\eta ,a_2\right)+4 \GG\left(1,a_3\right) \GG\left(a_1,a_2\right)+2 \GG\left(0,a_3\right) \Big(\GG\left(\eta ,a_2\right)\\
 &-\GG\left(a_1,a_2\right)-\GG\left(a_3,a_2\right)\Big)+4 \GG\left(0,a_1,a_2\right)+2 \GG\left(0,a_1 a_3,a_2\right)-4 \GG\left(\eta ,0,a_2\right)\\
 &+2 \GG\left(\eta ,a_1,a_2\right)-4 \GG\left(\eta ,a_3,a_2\right)+2 \GG\left(\eta ,a_1 a_3,a_2\right)+4 \GG\left(a_1,0,a_2\right)\\
 &-2 \GG\left(a_1,a_1 a_3,a_2\right)+4 \GG\left(a_3,0,a_2\right)-4 \GG\left(a_3,\eta ,a_2\right)-2 \GG\left(a_3,a_1 a_3,a_2\right)\Big) \GG\left(0,a_1\right)\\
 &-\GG\left(0,a_3\right) \GG\left(1,a_1\right)^2-2 \GG\left(1,a_1\right) \GG\left(\eta ,a_2\right)^2-2 \GG\left(1,a_1\right) \GG\left(a_3,a_2\right)^2\\
 &+2 \GG\left(1,a_1\right) \GG\left(a_1 a_3,a_2\right)^2-\GG\left(0,a_2\right)^2 \left(\GG\left(0,a_3\right)-2 \GG\left(1,a_1\right)\right)+\GG\left(0,a_3\right)^2 \GG\left(\eta ,a_2\right)\\
 &-2 \GG\left(0,a_3\right) \GG\left(1,a_1\right) \GG\left(\eta ,a_2\right)+4 \GG\left(1,a_1\right) \GG\left(1,a_3\right) \GG\left(\eta ,a_2\right)-\GG\left(0,a_3\right)^2 \GG\left(a_1,a_2\right)\\
 &-3 \GG\left(0,a_3\right)^2 \GG\left(a_3,a_2\right)+2 \GG\left(0,a_3\right) \GG\left(1,a_1\right) \GG\left(a_3,a_2\right)+2 \GG\left(0,a_3\right) \GG\left(1,a_1\right) \GG\left(a_1 a_3,a_2\right)\\
 &-4 \GG\left(1,a_1\right) \GG\left(1,a_3\right) \GG\left(a_1 a_3,a_2\right)+2 \GG\left(0,a_3\right) \GG\left(0,1,a_1\right)+2 \GG\left(0,a_3\right) \GG\left(0,a_1 a_3,a_2\right)\\
 &-4 \GG\left(1,a_1\right) \GG\left(0,a_1 a_3,a_2\right)+2 \GG\left(0,a_3\right) \GG\left(1,0,a_1\right)-2 \GG\left(\eta ,a_2\right) \GG\left(1,0,a_1\right)\\
 &+2 \GG\left(a_3,a_2\right) \GG\left(1,0,a_1\right)+\GG\left(0,a_2\right) \Big(\GG\left(0,a_3\right)^2-2 \GG\left(1,a_1\right) \GG\left(0,a_3\right)+2 \GG\left(1,a_1\right)^2\\
 &-4 \GG\left(0,1,a_1\right)-2 \GG\left(1,0,a_1\right)\Big)-2 \GG\left(\eta ,a_2\right) \GG\left(1,0,a_3\right)+2 \GG\left(a_1,a_2\right) \GG\left(1,0,a_3\right)\\
 &-2 \GG\left(0,a_3\right) \GG\left(\eta ,0,a_2\right)+4 \GG\left(1,a_1\right) \GG\left(\eta ,0,a_2\right)-4 \GG\left(1,a_1\right) \GG\left(\eta ,a_1,a_2\right)\\
 &+4 \GG\left(1,a_1\right) \GG\left(\eta ,a_3,a_2\right)+2 \GG\left(0,a_3\right) \GG\left(\eta ,a_1 a_3,a_2\right)+2 \GG\left(0,a_3\right) \GG\left(a_1,0,a_2\right)\\
 &-2 \GG\left(0,a_3\right) \GG\left(a_1,a_1 a_3,a_2\right)+4 \GG\left(0,a_3\right) \GG\left(a_3,0,a_2\right)-4 \GG\left(1,a_1\right) \GG\left(a_3,0,a_2\right)\\
 &+4 \GG\left(1,a_1\right) \GG\left(a_3,\eta ,a_2\right)-2 \GG\left(0,a_3\right) \GG\left(a_3,a_1 a_3,a_2\right)-4 \GG\left(1,a_1\right) \GG\left(a_1 a_3,0,a_2\right)\\
 &+4 \GG\left(1,a_1\right) \GG\left(a_1 a_3,a_1,a_2\right)-8 \zeta_3\Big]+\frac{1}{2} \zeta_2 \Big(5 \GG\left(\eta ,a_2\right)-\GG\left(0,a_1\right)\\
 &+5 \GG\left(0,a_2\right)-3 \GG\left(0,a_3\right)+\GG\left(1,a_1\right)-5 \GG\left(a_1,a_2\right)-5 \GG\left(a_3,a_2\right)\Big)\ .
 \end{align*}

\section{Conclusions}
\label{sec:conclusions}
In this paper we gave a detailed account of the calculation of different classes of two-loop Feynman integrals which evaluate to elliptic polylogarithms.
The main goal has been to show that the direct integration algorithms used to deal with 
polylogarithmic Feynman integrals
can be suitably generalised to include a considerably large class of elliptic Feynman integrals.
We considered in detail two different two-loop non-planar three-point functions appearing in the production of $t\bar{t}$ pairs in QCD and in the electroweak form factor, respectively. Subsequently, we computed the famous two-loop kite integral with three different internal masses. 
As it is well known, the kite integral inherits its elliptic nature from the two-loop massive sunrise subtopology.
The two families of three-point functions on the other hand are genuinely elliptic at the level of the top topology and in particular contain elliptic curves that are not directly related to the sunrise case.
In spite of their apparent diagrammtic similarity, these two families of three-point functions differ substantially. In fact, in the $t\bar{t}$ case 
the elliptic top-sector can be reduced to two independent master integrals (up to simpler polylogarithmic sub-topologies), 
while in the electroweak form factor case, the elliptic sector is reduced to three independent master integrals (again, up to subtopologies).

By direct integration over their Feynman parameter representation, we have shown that 
the complete set of master integrals for the three problems above can be consistently 
expressed in terms of elliptic polylogarithms. 
We have performed all computations in the Euclidean
region, such that all our results are real. In doing so, we described how to define our integrals consistently
depending on the location in the complex plane of the four branch points that define the elliptic curve.
As a crucial result of our calculations, we showed that for all examples considered it was always
possible to organise the master integrals of a given topology into a basis of \textsl{pure} building blocks. 
We stress here once more that by calling it a basis we mean that we can find a number of independent
pure building blocks which is equal to the number of master integrals in the problem under consideration.
This supports the conjecture according to which a basis of pure functions should always exist 
for master integrals that can be expressed in terms of MPLs only, and constitutes a strong hint towards
its generalisation to the elliptic case.

Finally,  there are strong indications that also many two-loop four-point Feynman integrals can be expressed in terms of the same
set of functions and that similar considerations on their transcendentality properties apply.  
We postpone the details of these calculations to future publications.

\section*{Acknowledgments}

We would like to thank the Mainz Institute for Theoretical Physics (MITP) in the context of the workshop ``High Time for Higher Orders: 
From Amplitudes to Phenomenology'' and the Galileo Galilei Institute in Florence in the context of the workshop ``Amplitudes in the LHC era''  for their hospitality and partial support during different phases of this work. This work was completed at the ETH Institute for
Theoretical Studies in the framework of the program "Periods, modular
forms and scattering amplitudes". We thank the institute for its
hospitality.
This research was supported by the ERC grant 637019 
``MathAm'', and the U.S. Department of Energy (DOE) under contract DE-AC02-76SF00515.

\appendix


\section{An example of analytic continuation}
\label{sec:top-empl}
In Section \ref{sec:ttbar-first-master}, we presented a result for the first master integral for the two-loop triangle integral relevant for top production in terms of eMPLs. The result shown in eq.~\eqref{eq:integral_final_v1} was however not well defined in the Euclidean region for which  $0 \leq a \leq \frac{1}{16}$  as there were poles on the integration contour. In this section, we show how to circumvent this problem by rewriting each $\mathcal{E}_4$ appearing in eq.~\eqref{eq:integral_final_v1} as a well defined combinations of eMPLs without poles on the integration contour.  This can be achieved through manipulations similar to how the integrand of eq.~\eqref{eq:last_int} was converted into eMPLs, i.e. by taking a derivative with respect to the kinematic variables $r_{+/-}$ and integrating back, thus ensuring the end-point of the contour for every eMPL to be $r_{+/-}$. In order to fix the imaginary parts of the eMPLs, we fix the imaginary parts of $r_{+/-}$ using a Feynman $i\varepsilon$ prescription which amounts to giving a small positive imaginary part to $q^2$ in eq.~\eqref{eq:adef}. Using eq.~\eqref{eq:rs} this gives in the Euclidean region
\begin{align}
\label{eq:ieps}
\text{Im}(a) > 0\,,\quad \text{Im}(\rmm) > 0\,,\quad\text{Im}(\rmp) < 0\ .
\end{align} 
We recall here that for the problem at hand $a$ and $r_{+/-}$ are not independent, see eq.~\eqref{eq:rs}. 
Nevertheless, we imagine to treat them as independent
variables in what follows.

The expression in eq.~\eqref{eq:integral_final_v1} contains eMPLs of length 3 and 4, which must be recast as eMPLs with endpoints $r_{+/-}$. Like for ordinary MPLs, this procedure consists of taking a derivative, thus generating lower-length eMPLs and integrating these back in the variable $r_{+/-}$. Therefore, one needs to work out first the length 1 examples in order to build the higher length iteratively. In the following we show examples of this procedure and in particular how to compute the boundary constants in some intricate cases.
\subsection{Length 1 example}
As a warm-up, consider the eMPL $\cEf{1}{\rmm}{1}$ which is simply a logarithm, $\log(1-1/\rmm)$. Taking a derivative with respect to $r_-$ we find
\begin{align}
\begin{split}
\frac{\partial}{\partial \rmm} \cEf{1}{\rmm}{1} \,=\, \frac{\partial}{\partial \rmm} \int_0^{1} \frac{dx}{x-\rmm} \,=\, \int_0^{1} \frac{dx}{(x-{\rmm})^2}\,=\,-\frac{1}{\rmm(1-\rmm)}\ .
\end{split}
\end{align}
Then, integrating back we find
\begin{align}
\begin{split}
\int^{\rmm}\!\!d\rmm' \frac{\partial}{\partial \rmm'} \cEf{1}{\rmm'}{1} &\,=\,-\int^{\rmm}\!\!d\rmm' \frac{1}{\rmm'(1-\rmm')}\\
 \,=\, -\log(\rmm) +\log(1-\rmm)+c&\,=\, -\cEf{1}{0}{\rmm}+\cEf{1}{1}{\rmm} + c\ .
\end{split}
\end{align}
The integration constant $c$ is then fixed by taking the limit $\rmm\to0$. In situations like the example above, there are logarithmic divergences in this limit due to the lower integration boundary at zero. As such, it is necessary to extract divergent logarithms in order to compute the boundary constant. For $\text{Im}(\rmm)>0$ we have for the left-hand-side
\begin{align}
\begin{split}
&\lim_{\rmm\to0}   \cEf{1}{\rmm}{1}\,=\,\lim_{\rmm\to0}   \log(1-1/\rmm)\\
\,=\,&\lim_{\rmm\to0}  \left(\log(1-\rmm)-\log(\rmm)-i \pi\right)\,=\,-\lim_{\rmm\to0}\cEf{1}{0}{\rmm}+i \pi\   ,
\end{split}
\end{align}
whereas for the right-hand-side we have
\begin{align}
\begin{split}
\lim_{\rmm\to0} \left(-\cEf{1}{0}{\rmm}+\cEf{1}{1}{\rmm} + c\right)\,=\,-\lim_{\rmm\to0}\cEf{1}{0}{\rmm}+ c \ ,
\end{split}
\end{align}
thus fixing $c=i\pi$.
\subsection{Length 2 example}
We now consider a more intricate example of length two and explicit dependence on the elliptic curve, namely $\cEf{0 & 1}{0 & \rmm}{1}$, in order to illustrate the typical steps involved in the analytic continuation of certain eMPLs. Note that due to the presence of the kernel $\Psi_{1}(\rmm, x)$, the integration on the real axis $0 < x < 1$ cannot be performed for $0 < \rmm < 1$ because the integrand has a pole on the contour. This is the case for the $t\bar{t}$ triangle in the Euclidean region. To avoid this problem, our goal is to rewrite $\cEf{0 & 1}{0 & \rmm}{1}$ as a combination of eMPLs of the form $\cEf{\cdots}{\cdots}{\rmm}$  such that there are no poles in the new contour $0<x<\rmm$. Similarly to the previous length-one example, this is achieved by differentiating the expression with respect to $\rmm$ and integrating back in terms of eMPLs with a different end point.
The first step is
\beq\bsp
\frac{\partial}{\partial \rmm} \cEf{0 & 1}{0 & \rmm}{1} \,&=\, \frac{\partial}{\partial \rmm} \int_0^{1} dx_1 \Psi_0(0,x_1) \int_0^{x_1} dx_2 \Psi_{1}(\rmm, x_2) \\
 \,&=\,  \frac{\partial}{\partial \rmm} \int_0^{1} dx_1\frac{c_4}{ \omega_1 y} \int_0^{x_1} \frac{dx_2}{x_2-1}\ ,
\esp\eeq
where we used the explicit expressions for the eMPLs kernels in eqs.~\eqref{eq:pure_psi0} and \eqref{eq:pure_psi1} and we omit the explicit dependence on the branch points of the elliptic curve,
\beq
\vec{b}^{\,(\mathrm{Euc})}\,=\,(0,b_-,b_+,1) \,,\quad b_\pm \,=\, \frac{1}{2}(1\pm\sqrt{1-16a})\ .
\eeq
Note that the ordering above differs from that of eq.~\eqref{eq:rootsttbar} as we are now in a configuration with $0<a<1/16$ where all roots are real and ordered.

Taking the derivative in $\rmm$ explicitly under the integral sign (recall the definition of $Z_4$ in eq.~\eqref{eq:Z4_def}) and integrating in $x_1$ and $x_2$, we find
\beq\bsp
\label{eq:ddr-lenght2}
\frac{\partial}{\partial \rmm} \cEf{0 & 1}{0 & \rmm}{1} \,&=\, \frac{\left(b_- -1\right)}{2 \omega _1 y_{\rmm}}\cEf{-1}{\rmm}{1}- \left(\frac{\left(b_--1\right) Z_4(\rmm)}{2 y_{\rmm}}+\frac{1}{\rmm}\right)\cEf{0}{0}{1}\ .
\esp\eeq
In order to simplify intermediate expressions, we use eq.~\eqref{eq:Gstar_ttbar} and the fact that for this particular case we have
\beq
Z_4(0) = Z_4(1) = 0\,.
\eeq
We are not yet ready to integrate back in $\rmm$ due to the presence of $ \cEf{-1}{\rmm}{1}$ which is not in the desired form $\cEf{\cdots}{\cdots}{\rmm}$. Therefore, we need to work separately on $ \cEf{-1}{\rmm}{1}$ and rewrite it first in a suitable form. We start in the same way, i.e.~by taking a derivative with respect to $\rmm$,
\beq\bsp
\label{eq:ddr-length1}
\frac{\partial}{\partial \rmm}  \cEf{-1}{\rmm}{1}\,=&\,\frac{\partial}{\partial \rmm}  \int_0^{ 1} dx  \Big[\frac{y_{\rmm}}{y (x-\rmm )} + Z_4(\rmm)\frac{c_4}{y} \Big]\ .
\esp\eeq
Once again, computing the derivative explicitly and integrating in $x$ we find that it vanishes and therefore $\cEf{-1}{\rmm}{1}$ is a constant,
\beq\bsp
\frac{\partial}{\partial \rmm}  \cEf{-1}{\rmm}{1}\,=&\,0 \quad \Rightarrow \quad \cEf{-1}{\rmm}{1} \,=\, c_1\ ,
\esp\eeq
In order to compute $c_1$ we consider the limit $\rmm\rightarrow 0$. We know that in this limit $Z_4(\rmm) \rightarrow 0$ and thus we are left with only the first term in the integrand~\eqref{eq:ddr-length1},
\beq\bsp
c_1\,=\,\lim_{\rmm\to 0} \cEf{-1}{\rmm}{1}\,&=\,\lim_{\rmm\to 0} \int_0^{1} dx \frac{y_{\rmm}}{(x-\rmm- i \varepsilon)y}\ ,
\esp\eeq
where we show explicit the Feynman $i\varepsilon$ prescription according to eq.~\eqref{eq:ieps}.
Changing variables to $t = x/\rmm$ and taking the limit $\rmm\rightarrow 0$, the square roots above become very simple,
\beq\bsp
\lim_{\rmm\to 0} y_{\rmm} \,=\, - i \pi \sqrt{\rmm b_- b_+} \,,\quad \lim_{\rmm\to 0} y \,=\, - i \pi \sqrt{t \, \rmm b_- b_+}\ .
\esp\eeq
With this we can obtain the result for the integration constant $c_1$,
\beq\bsp
c_1\,&=\,\lim_{\rmm\to 0} \int_0^{1/\rmm} dt \frac{1}{(t-1-i \varepsilon)\sqrt{t}}\,\sim\, \int_0^{\infty} dt \frac{1}{(t-1- i \varepsilon) \sqrt{t}}\,=\, - i \pi\ .
\esp\eeq
We are now ready to plug the result $\cEf{-1}{\rmm}{1}=-i\pi$ back into eq.~\eqref{eq:ddr-lenght2} and integrate in $\rmm$. We get
\beq\bsp
\label{eq:almost-final-l2}
\cEf{0 & 1}{0 & \rmm}{1} \,&=\, -\cEf{0}{0}{1} (\cEf{1}{\infty }{\rmm}+\cEf{1}{0}{\rmm}) + i \pi  \cEf{0}{0}{\rmm}+ c_2\ ,
\esp\eeq
where $c_2$ is another integration constant which can again be determined by studying the limit $\rmm\to0$.
The left-hand side of eq.~\eqref{eq:almost-final-l2} diverges logarithmic in the limit (recall that $\cEf{1}{0}{\rmm}$ is simply $\log(\rmm)$),
\beq\bsp
\label{eq:divrhs}
&\lim_{\rmm\to 0} \left[\cEf{0}{0}{1} (-\cEf{1}{\infty }{\rmm}-\cEf{1}{0}{\rmm}) + i \pi  \cEf{0}{0}{\rmm}+ c_2\right]\\
 \,=\, &-\cEf{0}{0}{1}\Big[ \lim_{\rmm\to 0} \cEf{1}{0}{\rmm}\Big] +c_2\ ,
\esp\eeq
whereas for the right-hand side we need to first unshuffle the logarithmic divergence,
\beq\bsp
\label{eq:divlhs}
\lim_{\rmm\to 0}  \cEf{0 & 1}{0 & \rmm}{1} \,&=\,\lim_{\rmm\to 0} \left[ \cEf{0}{0}{1}  \cEf{1}{\rmm}{1}  - \cEf{1 & 0}{\rmm & 0}{1} \right]\\
 \,&=\, i \pi  \cEf{0}{0}{1} - \cEf{1 & 0}{0 & 0}{1}- \cEf{0}{0}{1} \Big[ \lim_{\rmm\to 0} \cEf{1}{0}{\rmm}\Big] 
\esp\eeq
where we used the polylogarithmic relation
\beq\bsp
\cEf{1}{\rmm}{1} = \cEf{1}{1}{\rmm} - \cEf{1}{0}{\rmm}+i \pi \ .
\esp\eeq
Equating eqs.~\eqref{eq:divrhs} and \eqref{eq:divlhs} we find 
\beq 
c_2 = i \pi  \cEf{0}{0}{1}- \cEf{1 & 0}{0 & 0}{1}\ ,
\eeq
and thus we get our final expression valid for $0< a<1/16$,
\beq\bsp
\label{eq:final-l2}
\cEf{0 & 1}{0 & \rmm}{1} &= -\cEf{0}{0}{1} \big[\cEf{1}{\infty }{\rmm}+\cEf{1}{0}{\rmm}  - i \pi\big]  - \cEf{1 & 0}{0 & 0}{1}+ i \pi  \cEf{0}{0}{\rmm}\,.
\esp\eeq
Since both 0 and 1 are branch points of the integral, we have that certain eMPLs simplify using the definitions of the periods and quasi-periods. For example, we have
\beq
\cEf{0}{0}{1} \,=\, \frac{1}{2}\ .
\eeq
The procedure shown for this particular example can be generalised for arbitrary length and used to transform the complete expression into a result valid in the Euclidean region. For illustrative purposes, in the following section we show the first master integral completely transformed using these techniques.

\subsection{Result for top-production first master in the Euclidean region}
\label{sec:top-pure}
In the last section, we showed how the eMPLs expression for the first master integral for the top-production triangle given in eq.~\eqref{eq:integral_final_v1} can be turned into an expression valid in the Euclidean region. In this section, we show the result in terms of pure eMPLs, which is valid in the Euclidean region $0<a<1/16$ and is of uniform weight 4. The result is
\begin{align}
I=\frac{32\omega_1}{q^2(1+\sqrt{1-16 a})}[T_0(a)+3T_-(a)+5T_+(a)+\mathcal{O}(\epsilon)]\end{align}
where the individual building blocks depend on $r_+,\,r_-$ separately and are given by
\begin{align}
\begin{split}
T_a\,=\,&
-\cEf{0 & -1 & 1 & 1}{0 & \infty & 0 & 0}{1}- 
 \cEf{0 & -1 & 1 & 1}{0 & \infty & 0 & 1}{1}- 
 \cEf{0 & -1 & 1 & 1}{0 & \infty & 1 & 0}{1}-
 \cEf{0 & -1 & 1 & 1}{0 & \infty & 1 & 1}{1}+\\
 &
\log(a)\Big[ \cEf{0 & -1 & 1}{0 & \infty & 0}{1} + 
 \cEf{0 & -1 & 1}{0 & \infty & 1}{1}\Big]+ 
 \frac{1}{2} \cEf{0 & -1}{0 & \infty}{1}[\zeta_2 - \log^2(a)]\ ,
 \end{split}
\end{align}
\begin{align}
\begin{split}
T_-\,=\,&-\frac{3}{2} \zeta_2 \, \cEf{-1}{\infty}{\rmm}+ 
 \zeta_2 \cEf{-1 & 0}{\infty & 0}{\rmm}- 
 2 \cEf{-1 & -1}{\infty &\infty}{\rmm} \cEf{0 & -1}{0 & \infty}{1} \\
 & +
 \cEf{-1 & 0 & 1 & 1}{\infty & 0 & 0, 0}{\rmm}+ 
 \cEf{-1 & 0 & 1 & 1}{\infty & 0 & 0 & 1}{\rmm}- 
 \cEf{-1 & 0 & 1 & 1}{\infty & 0 & 1 & 0}{\rmm}- 
 \cEf{-1 & 0 & 1 & 1}{\infty & 0 & 1 & 1}{\rmm}\\
 &+ 
 \cEf{-1 & 1 & 0 & 1}{\infty & 0 & 0 & 1}{\rmm}- 
 \cEf{-1 & 1 & 0 & 1}{\infty & 1 & 0 & 0}{\rmm}+ 
 \cEf{1 & -1 & 0 & 1}{0 & \infty & 0 & 1}{\rmm}- 
 \cEf{1 & -1 & 0 & 1}{1 & \infty & 0 & 0}{\rmm}\\
 & - 
 \cEf{-1 & 0 & 1}{\infty & 0 & 1}{\rmm}\log(\rmm)+ 
 \cEf{-1 & 0 & 1}{\infty & 0 & 0}{\rmm}\log(1-\rmm)\ ,
\end{split}
\end{align}
\begin{align}
\begin{split}
T_+\,=\,&\frac{i \pi}{4}  \Big[\cEf{1& -1}{0& \infty}{\rmp} + 
    \cEf{1& -1}{1& \infty}{\rmp} - 
    4 \big(\cEf{1& -1& 0}{0& \infty & 0}{\rmp} + 
       \cEf{1& -1& 0}{1& \infty & 0}{\rmp}\big)\Big]
       \\& - 
 \cEf{1& -1& 0& 1}{0& \infty & 1& 0}{\rmp} 
 + 
 \cEf{1& -1& 0& 1}{0& \infty & 0& 1}{\rmp} - 
 \cEf{1& -1& 0& 1}{1& \infty & 1& 0}{\rmp} + 
 \cEf{1& -1& 0& 1}{1& \infty & 0& 1}{\rmp}\,.
\end{split}
\end{align}

\bibliography{bib}

\end{document}